\documentclass{aa}
\usepackage{txfonts}
\usepackage{graphicx}

\begin{document}

\newcommand{\be}{\begin{equation}}
\newcommand{\ee}{\end{equation}}
\newcommand{\bea}{\begin{eqnarray}}
\newcommand{\eea}{\end{eqnarray}}

\title{Simulated synchrotron emission from Pulsar Wind Nebulae}

\author{
        L. Del Zanna    \inst{1}
\and    D. Volpi        \inst{1} 
\and    E. Amato        \inst{2}
\and    N. Bucciantini  \inst{3} 
}

\offprints{Luca Del Zanna \\ \email{ldz@arcetri.astro.it}}

\institute{
Dipartimento di Astronomia e Scienza dello Spazio,
Universit\`a di Firenze, Largo E. Fermi 2, 50125 Firenze, Italy
\and
INAF - Osservatorio Astrofisico di Arcetri, Largo E. Fermi 5,
50125 Firenze, Italy
\and
Astronomy Department, University of California at Berkeley,
601 Campbell Hall, Berkeley, CA 94720-3411, USA
}

\date{Received ... Accepted ...}

\authorrunning{L.~Del Zanna et al.}

\abstract
{}
{A complete set of diagnostic tools aimed at producing synthetic synchrotron 
emissivity, polarization, and spectral index maps from relativistic MHD 
simulations is presented. As a first application we consider here the case 
of the emission from Pulsar Wind Nebulae (PWNe).}
{The proposed method is based on the addition, on top of the basic set of MHD 
equations, of an extra equation describing the evolution of the maximum energy
of the emitting particles. This equation takes into account adiabatic and 
synchrotron losses along streamlines for the distribution of emitting 
particles and its formulation is such that it is easily implemented in any 
numerical scheme for relativistic MHD.}
{Application to the axisymmetric simulations of PWNe, analogous to those 
described by Del Zanna et al. (2004, \aap, 421, 1063), allows direct 
comparison between the numerical results and 
observations of the inner structure of the Crab Nebula, and similar objects,
in the optical and X-ray bands. We are able to match most of the observed 
features typical of PWNe, like the equatorial torus and the polar jets, with
velocities in the correct range, as well as finer emission details, like arcs,
rings and the bright \emph{knot}, that turn out to arise mainly from
Doppler boosting effects. Spectral 
properties appear to be well reproduced too: detailed spectral index maps
are produced for the first time and show softening towards the PWN outer
borders, whereas spectral breaks appear in integrated spectra.
The emission details are found to strongly depend on both the average 
wind magnetization, here $\sigma_\mathrm{eff}\approx 0.02$, and on 
the magnetic field shape.} 
{Our method, in spite of its simplicity, provides a realistic modeling of 
synchrotron emission properties, and two-dimensional axisymmetric 
relativistic MHD simulations appear to be well suited to explain the main 
observational features of PWNe.}


\maketitle

\section{Introduction}
\label{sect:intro}

Non-thermal radiation, and synchrotron emission from relativistic
particles in magnetized plasmas in particular, is a ubiquitous
signature in high energy astrophysics. Strong shocks in relativistic flows 
are believed to be the natural site for the acceleration of particles
up to relativistic velocities. High energy electrons and positrons, then, 
efficiently emit synchrotron radiation while spiraling around magnetic 
field lines. For optically thin sources, the emission spectrum, 
typically a power law, is related
to the particles' distribution function at injection through well known
transport and radiation physics. However, the connection may be not
easy to establish quantitatively in the presence of multi-dimensional 
flow patterns, making it difficult to extract from observations direct 
information on the most basic physics of the sources.

Environments where synchrotron emission is encountered span from 
extragalactic sources, like cluster halos or Active Galactic Nuclei (AGNs), 
for example jets and lobes in radio-loud quasars, to the bulge and disk of 
our Galaxy, and galactic sources like Supernova Remnants (SNRs). 
Precisely to a subclass of the latter belongs
one of the most studied astrophysical sources: the Crab Nebula.
The Crab Nebula is the prototype of the so-called Pulsar Wind Nebulae (PWNe),
or plerions, and it is the object where synchrotron radiation was first
identified (Shklovsky \cite{shklovsky53}). The emission from PWNe displays
all the characteristic properties of synchrotron radiation from 
relativistic particles: a continuous, very broad-band spectrum,
extending from radio to X-rays and beyond (though Inverse Compton typically 
dominates at gamma-ray frequencies) and a high degree
of linear polarization; spectral indices, 
defined by $F_\nu\propto\nu^{-\alpha}$ where
$F_\nu$ is the net flux, range between 
$0\la\alpha\la 1.2$ and in a given source usually steepen with increasing 
frequency, 

While the theoretical modeling of the emitted radiation soon
reached a high level of sophistication (e.g. Kennel \& Coroniti 
\cite{kennel84b} for PWNe), what has been lacking until 
very recently is an appropriate modeling of the fluid and magnetic structure 
of the relativistic plasma responsible for the emission.
In the last few years, however, there has been an increasingly rapid growth
in accuracy and robustness of shock-capturing numerical schemes for 
relativistic hydrodynamics and MHD (see Mart\'i \& M\"uller \cite{marti03}
and Font \cite{font03} for periodically updated reviews), 
so that nowadays relativistic simulations are becoming a standard 
investigation tool for plasma astrophysics.
Among the others, we mention here the works on special relativistic MHD 
by Komissarov (\cite{komissarov99}) and Del Zanna et al. (\cite{delzanna03}), 
with the latter putting forward a simpler and yet powerful alternative
to characteristics-based algorithms. Based on these 
techniques, conservative shock-capturing schemes for MHD in General Relativity 
(GRMHD) have also been constructed (e.g. Gammie et al. \cite{gammie03};
Komissarov \cite{komissarov05}; Ant\'on et al. \cite{anton06}),
especially for studies of the accretion flows around Kerr black holes or
extraction of gravitational energy and subsequent jet acceleration. 
Recently, coupling of conservative schemes for GRMHD with solvers for
the Einstein equations has been achieved too (Duez et al. \cite{duez05}), 
for studies of binary neutron star mergers, core collapse in supernovae, 
black hole formation and gamma-ray burst generation.

The goal of the present paper is to provide the numerical community with
a simple computational tool that can be used to compute the brightness,
polarization and spectrum of the synchrotron light associated with
the plasma dynamics resulting from a given MHD simulation.
The method proposed is quite general and can be applied to any scheme 
for relativistic MHD (not necessarily in conservative form, either in 
full 3-D or under particular symmetries) and it just requires the numerical 
solution of an additional evolution equation, as we will discuss later.
As a first application of our techniques we study here in detail the
emission properties of PWNe, as they can be computed based on the recent 
axisymmetric model by Del Zanna et al. (\cite{delzanna04}). 

The interest of the scientific community for PWNe has recently received  
considerable impulse from the X-ray observations made by 
\emph{Chandra} and \emph{XMM-Newton}, showing that the Crab Nebula and 
other PWNe all present 
complex but similar inner structures: a higher emission torus in what is 
thought to be the equatorial plane of pulsar rotation, typically with 
brighter features like arcs or rings, a central \emph{knot}, 
and one or two opposite jets along the polar axis, 
apparently originating from very close to the pulsar position 
(Hester et al. \cite{hester95}, for \emph{Hubble} and \emph{ROSAT}
earlier observations; Weisskopf et al. \cite{weisskopf00}; 
Hester et al. \cite{hester02}; Helfand et al. \cite{helfand01}; 
Pavlov et al. \cite{pavlov01}; Gotthelf \& Wang \cite{gotthelf00}; 
Gaensler et al. \cite{gaensler01}, \cite{gaensler02};
Lu et al. \cite{lu02}; Camilo et al. \cite{camilo04}; 
Slane et al. \cite{slane04}; Romani et al. \cite{romani05}). 
This scenario is generally referred to as the \emph{jet-torus} structure.

While the presence of an equatorial torus of enhanced emission can
be explained by applying the standard 1-D relativistic MHD solutions
(Kennel \& Coroniti \cite{kennel84a}; Emmering \& Chevalier 
\cite{emmering87}) and assuming that the pulsar wind has a higher 
equatorial energy flux (Bogovalov \& Khangoulian \cite{bogovalov02}),
the origin of the polar jets has puzzled plasma physicists
until recently, given the known difficulties in obtaining 
self-collimation
via hoop stresses by the toroidal field in an ultra-relativistic MHD outflow
(Lyubarsky \& Eichler \cite{lyubarsky01} and references therein). 
However, an axisymmetric energy flux enhanced at the equator produces an 
oblate termination shock with a cusp-like shape at the poles. 
Moreover, if hoop stresses are at work in the mildly relativistic 
post-shock flow, jet collimation may occur on the axis at these cusps, 
thus giving the impression of an origin
inside the wind itself (Lyubarsky \cite{lyubarsky02}). This scenario 
has been recently confirmed by relativistic MHD numerical simulations
(Komissarov \& Lyubarsky \cite{komissarov03}, \cite{komissarov04};
Del Zanna et al. \cite{delzanna04}), which all show that the termination
shock really assumes the predicted shape, polar jets with velocities
in the observed range $0.5-0.8\,c$ are formed (provided the wind 
magnetization is high enough (approximately $1\%$), 
and the complex flow patterns which arise
near the termination shock are responsible for most of the observed
brighter features, basically just due to Doppler boosting effects 
(Komissarov \& Lyubarsky \cite{komissarov04}).

Compared to the quality of the simulation results for the flow structure,
synchrotron emission modeling has been so far rather crude. In the only 
attempt so far at computing synchrotron surface brightness maps on top of 
the flow structure obtained from 2-D MHD simulations 
(Komissarov \& Lyubarsky \cite{komissarov04})
radiative losses were approximated by an exponentially decaying factor,
with a spatially constant cooling time depending on the frequency 
of observation and an average magnetic field.
A power law distribution function for the emitting particles, normalized 
to the local thermal pressure to take into account variation 
of specific volume, was assumed in the emissivity. 
A more sophisticated method was proposed by Shibata et al. (\cite{shibata03}): 
here several particle energies are evolved independently in order to 
reconstruct the initial distribution function at any location and time. 
However, this method is suitable only for nearly steady-state solutions
where the individual paths (the fluid streamlines) can be unambiguously
followed. The method proposed here is a step forward concerning
the application to time-dependent numerical schemes: emitting particles
are continuously injected at the termination shock and then the full
particles' energy equation is integrated in time. However, differently
from the the above cited paper, just the \emph{maximum} energy attainable is
evolved, and it is taken as the upper cut-off in the local power law
distribution function that enters the emission coefficient.

Other than just synchrotron surface brightness, we also show here 
how to build polarization maps, namely maps of Stokes parameters,
degree of linear polarization, and position angle.
Such data are available for the Crab Nebula in radio (Wilson \cite{wilson72};
Velusamy \cite{velusamy85}) and optical (Schmidt et al. \cite{schmidt79};
Hickson \& van der Berg \cite{hickson90}; Michel et al. \cite{michel91}),
though high resolution data are still missing. X-ray polarization maps
are not available yet.
Synthetic maps have been reported in a preliminary work (Bucciantini et al.
\cite{bucciantini05b}), where optical maps have been produced for one
test case, here we discuss the dependence on the model parameters.
Finally, our method allows us to produce detailed spectral maps, which
can be compared to optical observations (e.g. V\'eron-Cetty \& Woltjer
\cite{veron-cetty93}) and to \emph{Chandra} and \emph{XMM-Newton} X-ray 
data (Weisskopf et al. \cite{weisskopf00}; Willingale et al. 
\cite{willingale01}; Mori et al. \cite{mori04}). Synthetic spectral
index maps can provide us with an additional diagnostic tool for the
inner magnetic structure of PWNe, since the spectral softening
due to synchrotron losses depends on the local magnetic field.
To our knowledge, this is the first time synthetic synchrotron spectral 
index maps based on data from numerical simulations have 
been produced.  

The paper is structured as follows: the proposed method is described in
Sect.~\ref{sect:sync}, the main PWN model assumptions are summarized in
Sect.~\ref{sect:model}, numerical results and synthetic synchrotron maps
are shown in Sect.~\ref{sect:results}, where also a comparison with 
observations is attempted. Finally, Sect.~\ref{sect:conclusions} is 
dedicated to the conclusions.

\section{Synthetic synchrotron emission recipes}
\label{sect:sync}

In the present section a method for producing synthetic synchrotron 
surface brightness and polarization maps, based on the results of numerical
simulations, will be derived and discussed.
The proposed method takes into account both adiabatic and
synchrotron losses in the evolution of the emitting particles advected
by the nebular flow, and it can be easily implemented in any numerical
scheme for relativistic MHD equations. The only assumption made here
is that the code provides the variables $\rho$ (proper mass density), 
$\vec{\varv}$ (flow bulk velocity), $p$ (proper thermal pressure),
and $\vec{B}$ (magnetic field) as functions of space, with or without
special symmetries reducing the dimensionality of the system, and time.
Lorentz transformations leading to relativistic corrections in the
emission properties, like Doppler shift and boosting or polarization 
angle swing, will also be fully considered. Therefore, in the following 
subsections, different notations will be employed where ambiguities may
occur: primed quantities will refer to the reference frame comoving with
the fluid, whereas standard notation will indicate quantities in the 
observer's frame.

\subsection{Distribution function evolution}
\label{sect:loss}

The observed spectra in astrophysical sources of non-thermal radiation
can be often approximated by power laws, with varying spectral index
$\alpha$ in different frequency regimes. It is therefore natural to
model the emitting particles' (typically electrons and positrons) distribution
function at injection locations as a combination of power laws as well.
Let then $\epsilon_0$ be the particle's energy (in units of the rest
mass energy $mc^2$) and consider a distribution function which is
isotropic in momentum and power law in energy, within a given energy range:
\be
\label{eq:f0}
f_0(\epsilon_0)=\frac{A}{4\pi}\epsilon_0^{-(2\alpha+1)},
~~~\epsilon_0^\mathrm{min}\leq\epsilon_0\leq\epsilon_0^\mathrm{max},
\ee
where from now on quantities labeled with the $0$ index will refer
to the injection region. The constant of proportionality 
\be
A=K_nn_0=K_p\frac{p_0}{mc^2}
\ee
can be determined by the definitions of the proper number density $n_0$ 
and thermal pressure $p_0$ in terms of $f_0(\epsilon_0)$, provided
$\epsilon_0^\mathrm{max}\gg\epsilon_0^\mathrm{min}$ (see Kennel \& Coroniti
\cite{kennel84b}). If $\alpha>1/2$, as it is the case in the following,
integration over momentum space and energy gives
\be
\left\{\begin{array}{l}
K_n = 2\alpha (\epsilon_0^\mathrm{min})^{2\alpha}, \\
K_p = 3 (2\alpha-1) (\epsilon_0^\mathrm{min})^{2\alpha-1},
\end{array}\right.
\ee
and from the above definitions we also find
\be
\epsilon_0^\mathrm{min}=3\left(\frac{2\alpha-1}{2\alpha}\right)
\frac{p_0}{n_0mc^2},
\ee
that can be determined by the Rankine-Hugoniot relations if the
injection location is a shock ($\epsilon_0^\mathrm{min}$ is of order
of the upstream wind Lorentz factor, $\sim 10^6$ for PWNe).

Time evolution of the single energies along post-shock streamlines
is governed by adiabatic and radiative losses. If $\mathrm{d}/\mathrm{d}
t^\prime=\gamma (\partial/\partial t + \vec{\varv}\cdot\nabla)$ is
the total time derivative in the comoving frame, the equation for
$\epsilon$ is
\be
\label{eq:epsilon}
\frac{\mathrm{d}}{\mathrm{d}\,t^\prime}\ln\epsilon =
\frac{\mathrm{d}}{\mathrm{d}\,t^\prime}\ln n^{1/3} + \frac{1}{\epsilon}
\left(\frac{\mathrm{d}\,\epsilon}{\mathrm{d}\,t^\prime}\right)_\mathrm{sync}
\ee
where synchrotron losses are
\be
\label{eq:loss}
\left(\frac{\mathrm{d}\,\epsilon}{\mathrm{d}\,t^\prime}\right)_\mathrm{sync}
\!\!\! = -\frac{4e^4}{9m^3c^5} {B^\prime}^2\epsilon^2,
\ee
in which we have averaged over all possible pitch angles with the 
local magnetic field that a particle may experience during its lifetime.
Based on the evolution of the single energies, we now seek an equation for
the evolution of the distribution function $f(\epsilon)$ along streamlines.
This problem was solved, for a stationary radial flow, by Kennel \& Coroniti
(\cite{kennel84b}), extended to the multi-dimensional case by
Begelman \& Li (\cite{begelman92}), and first applied to the results
of a time-dependent code by Bucciantini et al. (\cite{bucciantini05a}).
From the integration of Eq.~(\ref{eq:epsilon}) along streamlines it is
possible to write
\be
\frac{n^{1/3}}{\epsilon}-\frac{n^{1/3}_0}{\epsilon_0}=
\frac{n^{1/3}}{\epsilon_\infty},
\ee
where $\epsilon_\infty$ takes into account the integrated synchrotron
losses and corresponds to the remaining energy of a particle 
with energy $\epsilon_0\to\infty$ at the injection location, that is
to say the local maximum energy attainable. By using the above relation
it is possible to express $\epsilon$ as a function of $\epsilon_0$ and
$\epsilon_\infty$, which can be easily calculated for a stationary flow
everywhere in space. From the conservation of the particles' number 
$f(\epsilon)\mathrm{d}\epsilon/n=f_0(\epsilon_0)\mathrm{d}\epsilon_0/n_0$ 
along streamlines we are finally able to derive the distribution function as
\be
\label{eq:f_bl}
f(\epsilon)=\frac{K_p}{4\pi}\frac{p}{mc^2}\epsilon^{-(2\alpha+1)}
\left[\left(\frac{p}{p_0}\right)^{1/4}
\left(1-\frac{\epsilon}{\epsilon_\infty}\right)\right]^{2\alpha-1}
\!\!\!,\epsilon < \epsilon_\infty,
\ee
where we have used adiabaticity of the relativistically hot plasma:
$p\sim n^{4/3}$.

The method we propose in the time-dependent case, provided changes in
the global structure occur on timescales not shorter than the typical 
transport times, is to evolve
also the variable $\epsilon_\infty$ by adding an extra equation to
the relativistic MHD code. For a shock-capturing scheme the required
conservative form of Eq.~(\ref{eq:epsilon}) is found by combining it 
with the continuity equation for $\rho=nm$:
\be
\label{eq:infty}
\frac{\partial}{\partial t}(\gamma\rho \mathcal{E})+\nabla\cdot 
(\gamma\rho \mathcal{E}\vec{\varv})
=\frac{\rho \mathcal{E}}{\epsilon_\infty}
\left(\frac{\mathrm{d}\,\epsilon_\infty}
{\mathrm{d}\,t^\prime}\right)_\mathrm{sync},
\ee
with $\mathcal{E}=\epsilon_\infty/\rho^{1/3}$ and $\epsilon_\infty$ 
being a large enough number at injection, so as to mimic the theoretical 
behavior over the largest possible domain (differences due to 
$\epsilon_0<\infty$ should be confined to the close vicinities 
of the injection location, see Kennel \& Coroniti \cite{kennel84b}).

Once we have $\epsilon_\infty$ as a function of time and space, we can 
model the effects of adiabatic and synchrotron losses on the distribution
function by applying Eq.~(\ref{eq:f_bl}). For ease of implementation in 
a numerical code, however, a further approximation is finally needed.
Since $n_0$ and $p_0$ are difficult to trace back
starting from the position along the local streamline, in the 
multidimensional case one should also assume that their influence is
negligible. This occurs when two conditions are met: the dependency on
the term $(p/p_0)^{1/4}$ is weak and so is that on quantities involving the
temperature $\sim p_0/n_0$ at injection. Both conditions are easily
satisfied for $2\alpha-1\ga 0$, so that $K_n$ and $K_p$ are approximately
constant.

\subsection{The emission coefficient}
\label{sect:emis}

Consider an ultra-relativistic electron (or positron) with normalized
energy $\epsilon$, which spirals around the local magnetic field 
$\vec{B}^\prime$.
Its synchrotron spectral power (per unit frequency) is given by
\be
\label{eq:sp_power}
\mathcal{P}(\nu^\prime,\epsilon) = 
\frac{2e^4}{3m^2c^3} {B^\prime_\perp}^2\epsilon^2 
\mathcal{S}(\nu^\prime,\nu^\prime_c),
\ee
where $B^\prime_\perp$ is the field component normal to the particle's
velocity, the spectral density is 
\be
\mathcal{S}(\nu^\prime,\nu^\prime_c) = \frac{9\sqrt{3}}{8\pi\nu^\prime_c}
F\left(\frac{\nu^\prime}{\nu^\prime_c}\right),
\ee
and the function $F(x)$ (see Rybicki \& Lightman \cite{rybicki79} for its 
properties) is such that emission peaks around the critical frequency
\be
\nu^\prime_c(\epsilon) = \frac{3e}{4\pi mc}\,B^\prime_\perp\epsilon^2.
\ee
The local emission coefficient $j_\nu^\prime$ (radiated power per unit 
frequency, volume and solid angle), for given observation frequency 
$\nu^\prime$ and direction $\mathbf{n}^\prime$, is obtained by integrating 
the product of Eqs.~(\ref{eq:sp_power}) and (\ref{eq:f_bl}) over the whole 
range of local particle energies:
\be
\label{eq:j1}
j^\prime_\nu(\nu^\prime,\vec{n}^\prime)= \!\!
\int \mathcal{P} (\nu^\prime,\epsilon)f(\epsilon)\mathrm{d}\epsilon.
\ee
Note that, since radiation from ultra-relativistic 
particles is strongly beamed in their instantaneous direction of motion 
(within a cone of angle $1/\epsilon\ll 1$), only those particles with 
pitch angle coinciding with the angle between
$\mathbf{B}^\prime$ and $\mathbf{n}^\prime$ (thus $B^\prime_\perp
=|\vec{B}^\prime\times\vec{n}^\prime|$) contribute to the emission 
along the line of sight in Eq.~(\ref{eq:j1}).

The above integral can be calculated by replacing the energy with the
variable $x=\nu^\prime/\nu_c^\prime\propto\epsilon^{-2}$. In order to
take advantage of the known properties of the function $F(x)$ we
further simplify Eq.~(\ref{eq:f_bl}) to a standard power law form
\be
\label{eq:f}
f(\epsilon)=\frac{K_p}{4\pi}\frac{p}{mc^2}\epsilon^{-(2\alpha+1)}, ~~~
\epsilon < \epsilon_\infty,
\ee
in which we have assumed that $K_p$ is constant, the dependency on the term
$(p/p_0)^{1/4}$ is weak, and synchrotron burn-off can be modeled with a 
sharp cut-off at $\epsilon_\infty$. This last additional assumption is
again a good approximation provided $2\alpha-1\ga 0$. The result is
\be
\label{eq:j2}
j^\prime_\nu(\nu^\prime,\vec{n}^\prime)= 
C p |\vec{B}^\prime\times\vec{n}^\prime|^{\alpha +1}{\nu^\prime}^{-\alpha},
\ee
where all spatially independent terms have been gathered in the constant
\be
C \! =\! \frac{\sqrt{3}}{16\pi}
\frac{\alpha \! + \! 5/3}{\alpha + 1}
\Gamma\left(\frac{\alpha \! + \! 5/3}{2}\right)
\Gamma\left(\frac{\alpha \! + \! 1/3}{2}\right)
\frac{e^3}{m^2c^4}\left(\frac{3e}{2\pi mc}\right)^\alpha\! K_p,
\ee
in which the property $\Gamma(\xi+1)=\xi\Gamma(\xi)$ of the Euler gamma 
function $\Gamma(\xi)$ has been used.

In order to obtain the emissivity in the observer's fixed frame 
of reference, relativistic corrections must be taken into account. 
It is well known that both the frequency and the emission coefficient
itself transform via the Doppler boosting factor
\be
D=\frac{1}{\gamma \,(1-\vec{\beta}\cdot\vec{n})},
\ee
as $\nu=D\nu^\prime$ and $j_{\nu}=D^2 j_{\nu}^\prime$, where $\vec{\beta}
=\vec{\varv}/c$. The resulting emission coefficient is finally written as
\be
\label{eq:j}
j_{\nu}(\nu,\vec{n}) = \left\{\begin{array}{ll}
C p \,|\vec{B}^\prime\times\vec{n}^\prime|^{\alpha +1}
D^{\alpha +2}\nu^{-\alpha}, & \nu_\infty\ge\nu, \\
 & \\
0, & \nu_\infty < \nu. \end{array}\right.
\ee
The cut-off frequency for synchrotron burn-off is
\be
\label{eq:nu_infty}
\nu_\infty\equiv D \nu_c^\prime (\epsilon_\infty) =
D \frac{3e}{4\pi mc}\,|\vec{B}^\prime\times\vec{n}^\prime|
\epsilon_\infty^2,
\ee
where Doppler shift has been considered and where $\epsilon_\infty$ is
provided directly by the numerical scheme via integration of 
Eq.~(\ref{eq:infty}).

The terms $B^\prime$ in Eq.~(\ref{eq:loss}), needed to evolve
$\epsilon_\infty$, and
$|\vec{B}^\prime\times\vec{n}^\prime|$ in Eqs.~(\ref{eq:j},\ref{eq:nu_infty}),
are defined through quantities in the comoving frame. 
Application of the composition rule for relativistic velocities 
to the observer direction versor $\vec{n}$ yields
\be
\label{eq:n^prime}
\vec{n}^\prime=D\left[\vec{n}+\left(\frac{\gamma^2}{\gamma+1}
\vec{\beta}\cdot\vec{n}-\gamma\right)\vec{\beta}\right],
\ee
while Lorentz transformations of Maxwell's equations in the (ideal) MHD case
$\vec{E}+\vec{\beta}\times\vec{B}=0$ provide
\be
\label{eq:B^prime}
\vec{B}^\prime=\frac{1}{\gamma}\left[\vec{B}+\frac{\gamma^2}{\gamma+1}
(\vec{\beta}\cdot\vec{B})\vec{\beta}\right].
\ee
After some straightforward algebra, the required quantities can be written as
\be
B^\prime=\frac{1}{\gamma}\sqrt{B^2+\gamma^2(\vec{\beta}\cdot\vec{B})^2},
\ee
and
\be
\label{eq:b_times_n}
|\vec{B}^\prime\times\vec{n}^\prime|=\frac{1}{\gamma}
\sqrt{B^2-D^2(\vec{B}\cdot\vec{n})^2+2\gamma D(\vec{B}\cdot\vec{n})
(\vec{\beta}\cdot\vec{B})},
\ee
in which everything on the right-hand-side is now measured in 
the observer's frame.

\subsection{Surface brightness and polarization maps}
\label{sect:maps}

To compare the synthetic synchrotron emission obtained by numerical
simulations with real images, the first step certainly requires the
production of surface brightness maps. Let us consider a Cartesian
observer's reference frame in which $X$ lies along the line of sight
$\vec{n}$ and $Y$ and $Z$ are in the plane of the sky (say with $Z$
indicating the North). By switching from the local coordinate system,
which may present special symmetries, to the observer's system $(X,Y,Z)$
(taking into account the angles of inclination of the emitting object),
surface brightness (or specific intensity) maps are obtained as
\be
\label{eq:I}
I_\nu (\nu,Y,Z) = \int_{-\infty}^{\infty}j_\nu (\nu,X,Y,Z)\,\mathrm{d}X.
\ee

A powerful additional diagnostic tool may be provided by polarization 
measurements, where available. Synchrotron emission from relativistic 
particles is known to be linearly polarized, with a high degree of 
polarization, and Stokes parameters $U_\nu$ and $Q_\nu$ ($V_\nu=0$) can be 
calculated from numerical simulations by measuring the local polarization 
position angle $\chi$, that is the angle of the emitted electric field vector
$\vec{e}$ in the plane of the sky, measured clockwise from the $Z$ axis. 
Synthetic maps of the Stokes parameters are thus built as
\be
\label{eq:Q}
Q_\nu (\nu,Y,Z) = \frac{\alpha+1}{\alpha+5/3}\int_{-\infty}^{\infty}
j_\nu (\nu,X,Y,Z)\cos 2\chi\,\mathrm{d}X,
\ee
\be
\label{eq:U}
U_\nu (\nu,Y,Z) = \frac{\alpha+1}{\alpha+5/3}\int_{-\infty}^{\infty}
j_\nu (\nu,X,Y,Z)\sin 2\chi\,\mathrm{d}X,
\ee
where the additional factor comes from the intrinsic properties of
synchrotron emission (see Rybicki \& Lightman \cite{rybicki79}).
Related to the Stokes parameters above is the degree of (linear)
polarization, or polarization fraction, defined as
\be
\Pi_\nu=\frac{\sqrt{Q_\nu^2+U_\nu^2}}{I_\nu}.
\ee

In order to calculate $\chi$, relativistic effects like position
angle swing must be taken into account (Blandford \& K\"onigl 
\cite{blandford79}; Bj\"ornsson \cite{bjornsson82}; Lyutikov et al. 
\cite{lyutikov03}). 
In the comoving frame, the emitted electric field of the linearly
polarized radiation is normal to both the local magnetic field
$\vec{B}^\prime$ and the line of sight $\vec{n}^\prime$, thus
$\vec{e^\prime}\propto\vec{n}^\prime\times\vec{B}^\prime$, and
the radiated magnetic field is therefore 
$\vec{b^\prime}=\vec{n}^\prime\times\vec{e}^\prime$.
Lorentz transformations give the electric field in the observer frame as
\be
\vec{e}=\gamma\left[\vec{e^\prime}-\frac{\gamma}{\gamma+1}
(\vec{e^\prime}\cdot\vec{\beta})\vec{\beta}-
\vec{\beta}\times\vec{b}^\prime\right],
\ee
and by using Eqs.~(\ref{eq:n^prime}) and (\ref{eq:B^prime}) we can
express the required electromagnetic fields in terms of quantities 
defined in the observer frame.
After some calculations, the simple formula derived by
Lyutikov et al. (\cite{lyutikov03}) is retrieved:
\be
\vec{e}\propto\vec{n}\times\vec{q},~~~
\vec{q}=\vec{B}+\vec{n}\times (\vec{\beta}\times\vec{B}).
\ee
Notice that $\vec{e}$ lies in the plane of the sky, as expected,
and that $\vec{q}$ correctly reduces to $\vec{B}$ for non-relativistic
fluid bulk motions. We are now ready to derive the position angle $\chi$
in terms of the components of $\vec{q}$ in the plane of the sky, which are
\be
q_Y=(1-\beta_X)B_Y+\beta_Y B_X,~~~
q_Z=(1-\beta_X)B_Z+\beta_Z B_X.
\ee
The functions of the position angle appearing in Eqs.~(\ref{eq:Q},\ref{eq:U})
may be written in terms of these as:
\be
\cos 2\chi=\frac{q^2_Y-q^2_Z}{q^2_Y+q^2_Z},~~~
\sin 2\chi=-\frac{2q_Yq_Z}{q^2_Y+q^2_Z}.
\ee

\subsection{Spectral index maps and integrated spectra}
\label{sect:spectra}

In order to complete our set of diagnostic tools based on the
synthetic synchrotron emission obtained from numerical simulations,
we now turn to spectral properties.
An interesting possibility is provided by maps of spectral index.
This quantity is defined as
\be
\label{eq:index}
\alpha_\nu(\nu_1,\nu_2,Y,Z)=
-\frac{\log [I_\nu(\nu_2,Y,Z)/I_\nu(\nu_1,Y,Z)]}{\log(\nu_2/\nu_1)},
\ee
for any given interval in frequency $[\nu_1,\nu_2]$. In general,
due to synchrotron radiative losses, this quantity is expected to
increase, thus the spectrum is expected to steepen, when moving
away from the source of ultra-relativistic particles (spectral softening).

Finally, integrated spectra (or net fluxes) are obtained by averaging the 
specific intensity over the solid angle covered by the source in 
the plane of the sky, which may be calculated as
\be
\label{eq:netflux}
F_\nu(\nu)=\frac{1}{d^2}\int\!\!\int I_\nu(\nu,Y,Z)\,\mathrm{d}Y\mathrm{d}Z,
\ee
where $d$ is the distance of the astronomical object under consideration.

Notice that both the above quantities related to spectral features have a 
meaning only if synchrotron burn-off is modeled in some way by the numerical
simulation.

\section{Pulsar wind model and initial settings}
\label{sect:model}

As anticipated in the Introduction, let us now apply our set of
diagnostic tools to the modeling of the inner structure of PWNe.
Axisymmetric 2-D simulations
of the interaction of an ultra-relativistic magnetized pulsar wind 
with supernova ejecta expanding in the static interstellar medium
were described in Del Zanna et al. (\cite{delzanna04}), to which
the reader is referred for all the modeling details and for the 
results. Here we will calculate the synchrotron emission from the
results of new simulations, obtained by using very similar settings.
The numerical code employed is described in Del Zanna \& Bucciantini
(\cite{delzanna02}), Del Zanna et al. (\cite{delzanna03}), and
Londrillo \& Del Zanna (\cite{londrillo04}). It is a shock-capturing
code solving the ideal relativistic MHD equations in conservative 
form, to which Eq.~(\ref{eq:infty}) has been added. Note that the effects
of synchrotron radiative losses on the fluid total energy evolution 
are not taken into account. Radiative losses would become
important if the particles radiated away a considerable fraction of the
pulsar energy input to the nebula. However, the radiated power is estimated
to be of order $10\%$ of the instantaneous pulsar input in the case of the
Crab Nebula, and generally much smaller than this in the case of other
PWNe, so the condition of adiabaticity should generally be considered as
a good approximation.

Let us briefly summarize the initial conditions and the pulsar wind
model in particular. The numerical box is made up of 400 cells in
the radial direction, between $r=0.1$ ly and $r=10$ ly (a logarithmic
grid with 200 cells per decade is used to enhance resolution in the
inner region), and 100 cells in the latitudinal direction, 
with polar angle $\theta$ defined in the first quadrant only (symmetry 
conditions are applied at the equator $\theta=\pi/2$). 
Within an arbitrary radius of 0.2 ly we
impose initially the wind conditions, while the expanding ejecta,
with linearly increasing radial velocity, occupy the region up to 1 ly.
The (constant) wind luminosity is taken to be 
$L_0=5\times 10^{38}$~erg~s$^{-1}$,
the energy of the supernova explosion turned into kinetic energy
of the ejecta $E_\mathrm{ej}=10^{51}$~erg, and the total mass of
the ejected material $M_\mathrm{ej}=3 M_{\sun}$.
For similar settings and a description on the (1-D) evolution
of the PWN / SNR system see van der Swaluw et al. (\cite{vanderswaluw01})
and Bucciantini et al. (\cite{bucciantini03}). The chosen values were seen
to provide an overall evolution compatible with that of the Crab Nebula. 
Furthermore, the magnetic field $\vec{B}$ is assumed to retain just the 
toroidal component, as in Kennel \& Coroniti (\cite{kennel84a}), while
the velocity $\vec{\varv}$ is assumed to be always poloidal, so that
some of the relativistic corrections in Sect.~\ref{sect:emis} are
simplified since $\vec{\beta}\cdot\vec{B}=0$.

The latitudinal dependence of the wind energy flux is provided 
by the following choice for the wind bulk Lorentz factor 
\be
\label{eq:lorentz}
\gamma (\theta)=\gamma_0 \left[\,\alpha_0+(1-\alpha_0)\sin^2\theta\,\right],
\ee
where $\gamma_0=100$ is the equatorial value and $\alpha_0=0.1$ the
anisotropy parameter. Velocity is assumed to be in the radial direction
alone at $t=0$. The residual magnetic field at large distances from
the pulsar's light cylinder is thought to be mainly toroidal with a
dependence $B\sim\sin\theta/r$ (\emph{split monopole} models: Michel
\cite{michel73}) inside the wind.
By prescribing such a field and an isotropic mass flux, the energy
flux for a cold ultra-relativistic wind may be then written as
\be
\label{eq:flux}
F(r,\theta)=F_0 \left(\frac{r_0}{r}\right)^2
\left[\,\alpha_0+(1-\alpha_0 + \sigma) \sin^2\theta\,\right],
\ee
where $\sigma$ is the wind magnetization parameter, namely the ratio
of magnetic to kinetic energy fluxes computed at the equator
(Kennel \& Coroniti \cite{kennel84a}), and $F_0$ is 
the equatorial kinetic energy flux at the reference radius $r_0$, to be 
derived in terms  of $L_0$, $\alpha_0$, and $\sigma$ by integrating 
Eq.~(\ref{eq:flux}) over a spherical surface of radius $r$.

The value of $\sigma$ is typically expected to be a small number in order
to match the predicted and observed nebular emission
(Kennel \& Coroniti \cite{kennel84b}: $\sigma=3\times 10^{-3}$
for the Crab Nebula), thus 
one is left with the problem of how to convert a Poynting flux dominated 
outflow near the pulsar to a kinetically dominated wind before the termination
shock (the so-called \emph{sigma paradox}) (see e.g. Arons \cite{arons04}).
One of the solutions proposed
invokes the presence a \emph{striped wind} region around the equator,
due to the inclination of the pulsar's magnetosphere with respect to the 
rotation axis, where dissipation in the resulting current sheets may form
a neutral (unmagnetized) equatorial region (Coroniti \cite{coroniti90}). 
We model here such a situation by prescribing the wind toroidal field as 
\be
\label{eq:torfield}
B(r,\theta)=B_0 \left(\frac{r_0}{r}\right)\,\sin \theta \,\,
\tanh \left[b\left(\frac{\pi}{2}-\theta\right)\right],
\ee
where we set $B_0=\sqrt{4\pi\sigma F_0/c}$ and where
the last term mimics the conversion of magnetic to kinetic energy in the 
striped wind. The parameter $b$ controls the angular size of the zone 
of lower magnetization: a large $b$ indicates a narrow striped wind region,
and the \emph{ideal} limit is recovered by letting $b\to\infty$.
Here we assume that Eq.~(\ref{eq:flux}) still holds even for 
a finite $b$.
Finally, the density $\rho$, decreasing as $r^{-2}$, can be derived 
from the above equations and will depend on $\sigma$ and $b$ too, 
since it is supposed to account also for the conversion of magnetic 
to kinetic energy in the striped wind region (note that the mass 
flux is actually isotropic only in the ideal limit $b\to\infty$).

\begin{figure}
\resizebox{\hsize}{!}{\includegraphics{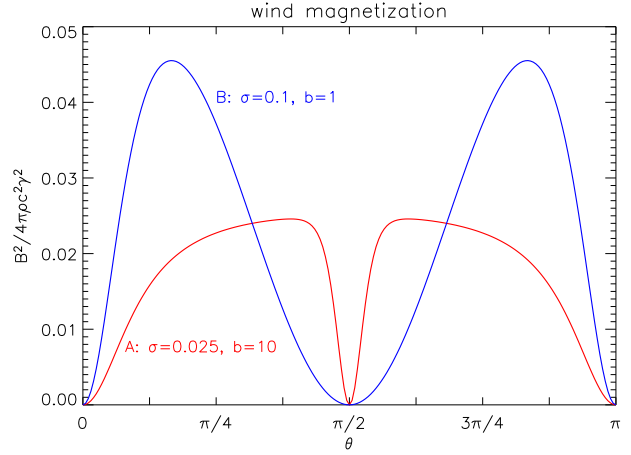}}
\caption{The wind magnetization $B^2/4\pi\rho c^2\gamma^2$ as a function
of the polar angle $\theta$, at any radius  $r$, for the two runs
defined in Eq.~(\ref{runs}).
In both cases the averaged value is approximately the same:
$\sigma_\mathrm{eff}\approx 0.02$.}
\label{fig:magnetization}
\end{figure}

As our main goal here is the comparison with observations of the
Crab Nebula, the temporal evolution of the PWN / SNR interaction
will not be followed in any detail. On the contrary, we will focus
our attention on the synchrotron emission properties of the simulated
PWN at a time corresponding to the age of the Crab, $t\approx 1000$~yr,
which is still in the free expansion phase (Bucciantini et al. 
\cite{bucciantini03}). As shown by Del Zanna et al. (\cite{delzanna04}), 
equatorial flows and polar jets with velocities in the range of the
observed values are found for an \emph{effective} magnetization 
of the wind plasma $\sigma_\mathrm{eff}\ga 0.01$, 
where $\sigma_\mathrm{eff}$ is obtained by averaging over polar angle 
(the magnetization in the wind region does not depend on $r$).
Thus, here we will consider two different runs, A and B, with approximately
the same $\sigma_\mathrm{eff}\approx 0.02$ and very different magnetic fields:
\be
\label{runs}
\left\{\begin{array}{lll}
\mathrm{run\,\,A:} & \sigma=0.025, & b=10, \\
\mathrm{run\,\,B:} & \sigma=0.1,   & b=1.
\end{array}\right.
\ee
The wind magnetization $B^2/4\pi\rho c^2\gamma^2$ for the two runs 
as a function of the polar angle $\theta$ is shown in 
Fig.~\ref{fig:magnetization}.
Notice the difference in the striped wind region width around the equator: 
run A refers to a nearly ideal case, as those discussed by
Del Zanna et al. (\cite{delzanna04}), while in run B the striped wind
region is very wide.

\section{Simulation results}
\label{sect:results}

\subsection{Flow structure}

\begin{figure*}
\centerline{\resizebox{1.05\hsize}{!}{
\includegraphics{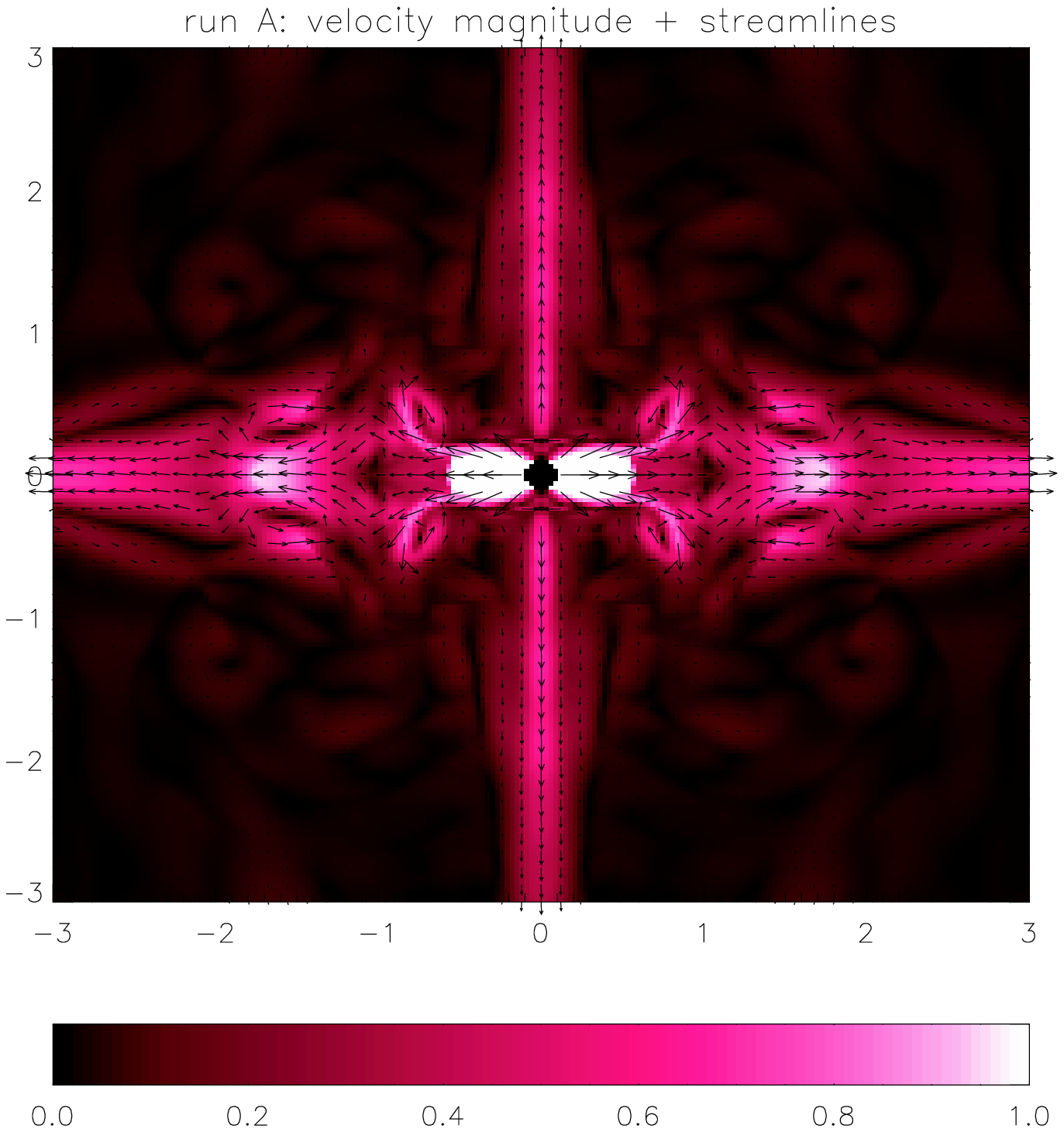}
\includegraphics{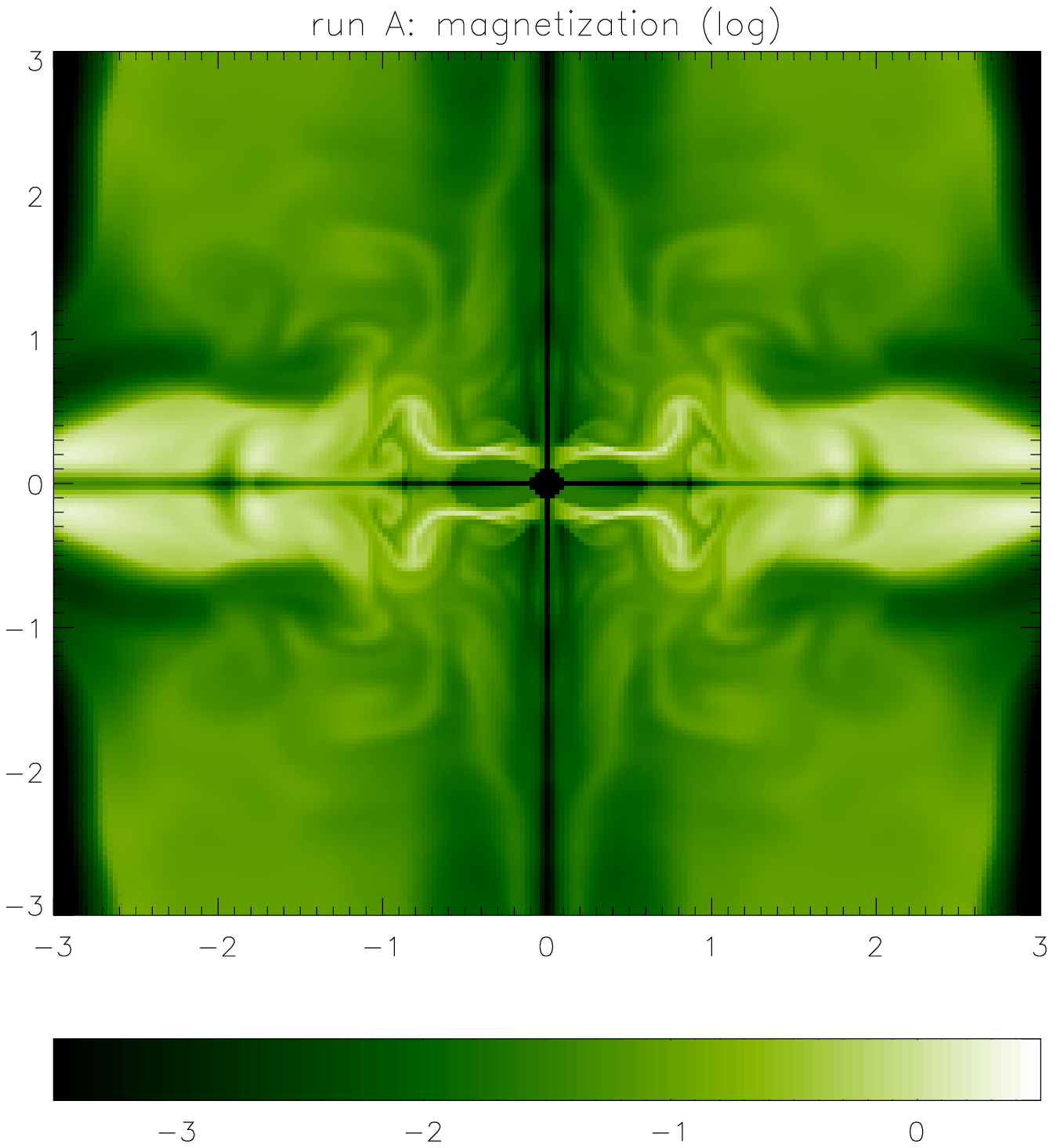}
\includegraphics{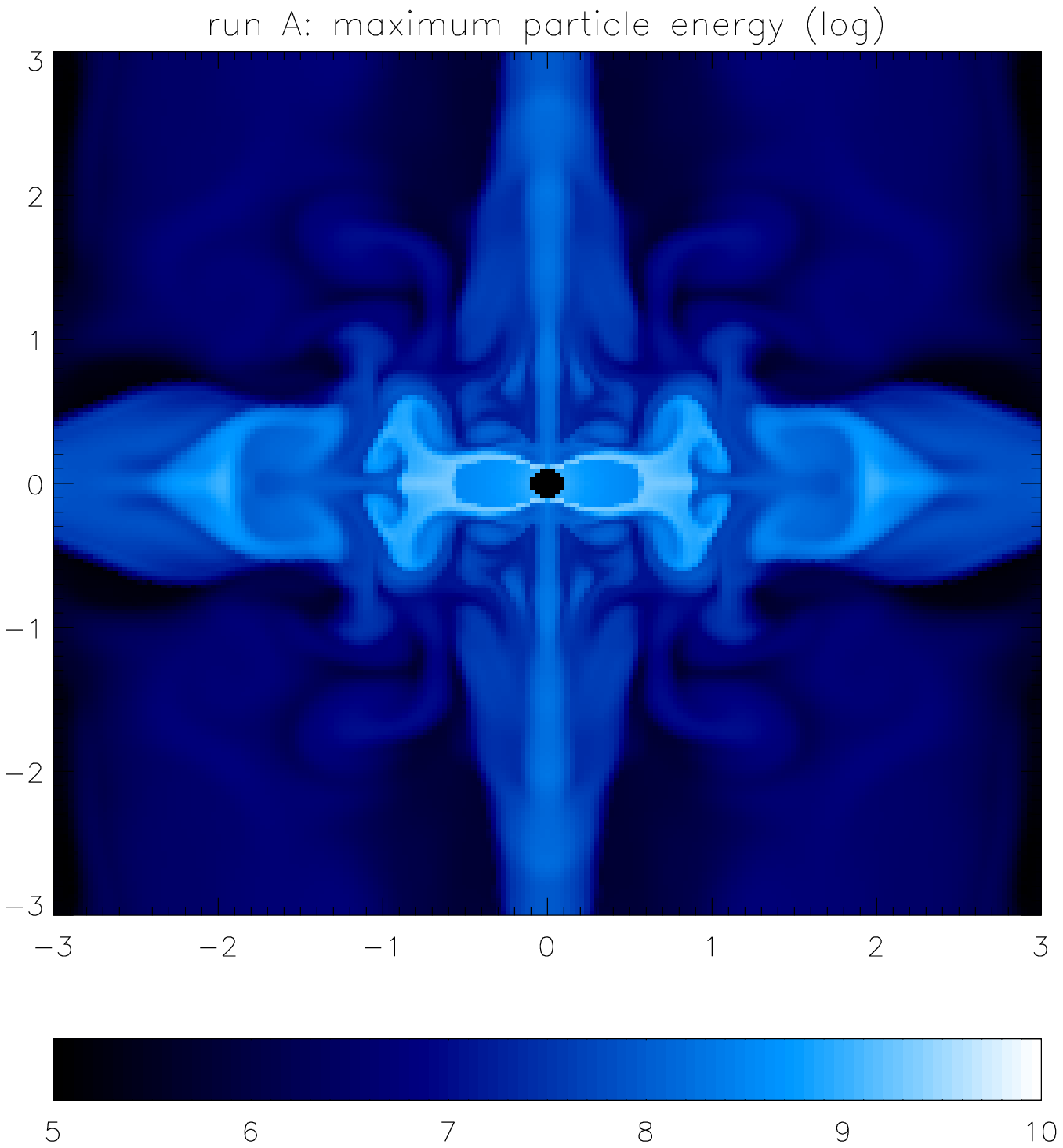}
}}
\centerline{\resizebox{1.05\hsize}{!}{
\includegraphics{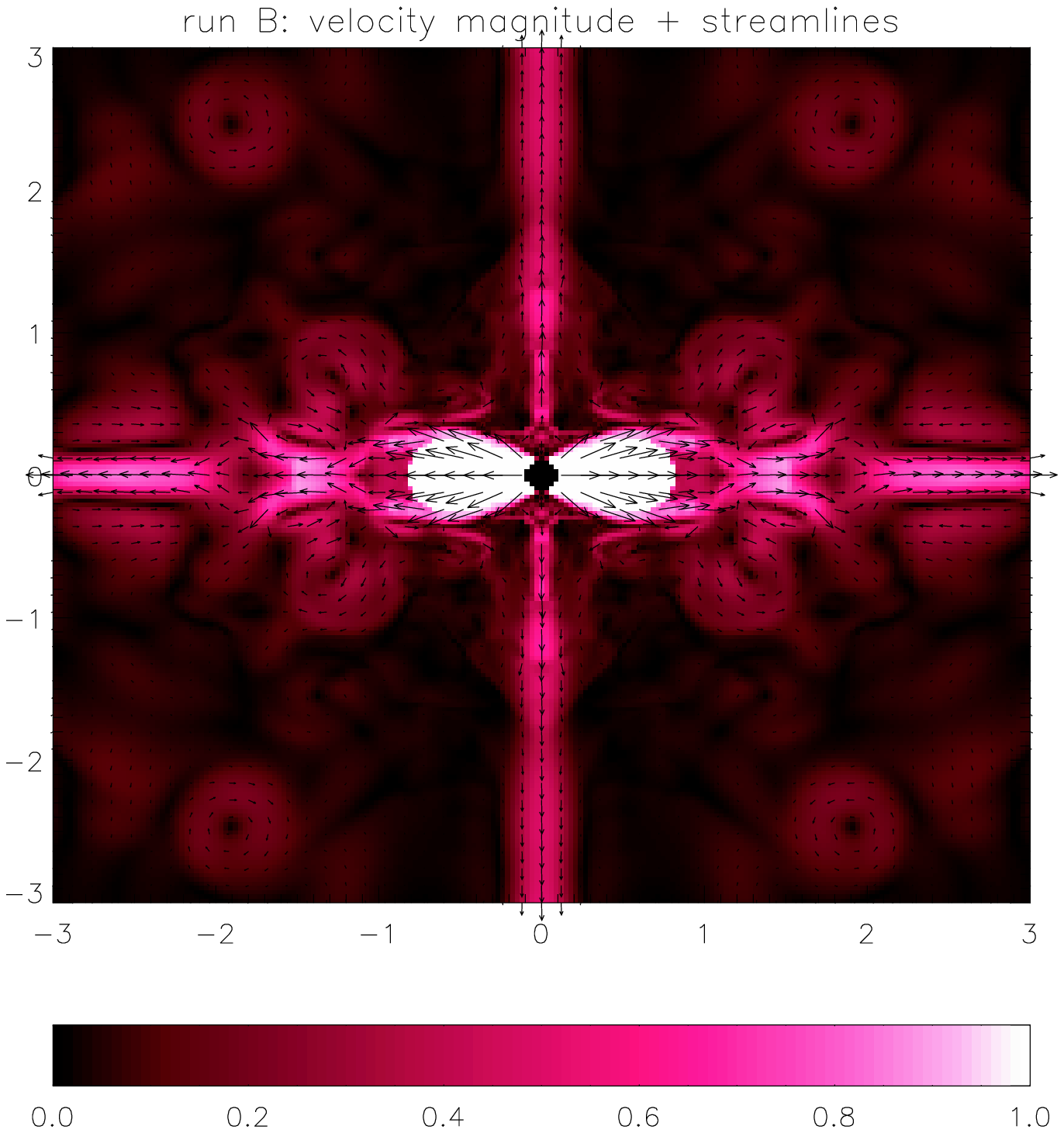}
\includegraphics{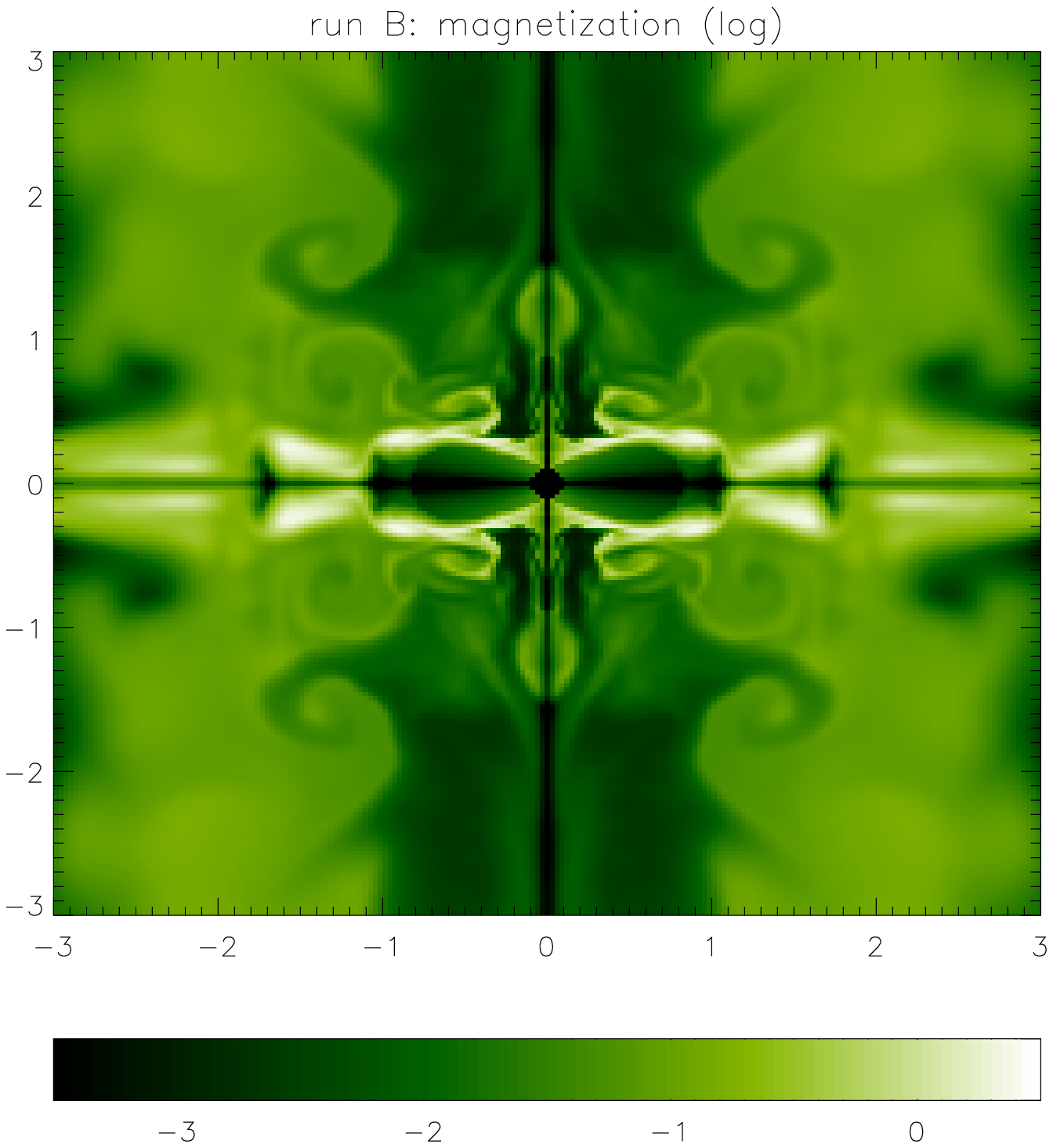}
\includegraphics{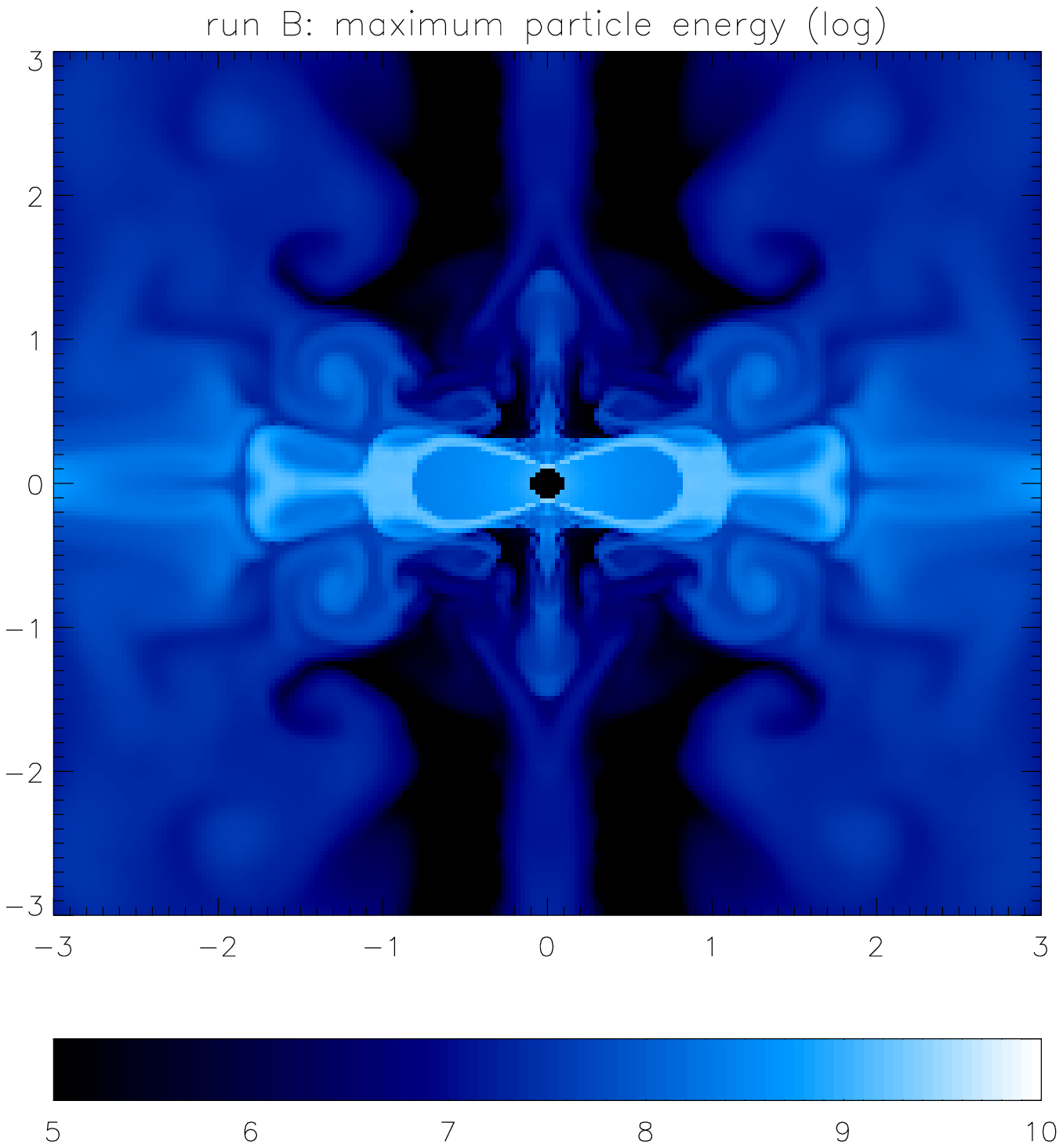}
}}
\caption{
Velocity magnitude (in units of $c$) and streamlines (left panels), 
magnetization (center panels), and maximum particle energy $\epsilon_\infty$ 
(in units of $m_ec^2$, right panels) 
in the inner region of the simulated PWN, for run A (upper row) 
and run B (lower row). Distances from the central pulsar
are reported on the axes, expresses in light year (ly) units. 
}
\label{fig:flow}
\end{figure*}

Before studying the synchrotron radiation maps, let us briefly comment
on the flow structure and related properties for the two different runs,
focusing on the quantities that are relevant for computing the emission.
In Fig.~\ref{fig:flow} we show the velocity field (velocity magnitude and
streamlines), the magnetization $B^2/4\pi w\gamma^2$ ($w=\rho c^2+4p$
is the enthalpy, dominated by the kinetic component in the pulsar wind
and by the thermal one inside the nebula), and the maximum particle 
energy $\epsilon_\infty$. The latter is evolved according to 
Eq.~(\ref{eq:infty}) and its value at injection is $10^9$. Initialization
of $\epsilon_\infty$ occurs, like for all other quantities,  
in the wind region, so that the value of $10^9$ is further enhanced 
at the crossing 
of the termination shock, due to the conservation of the quantity 
$\mathcal{E}=\epsilon_\infty/\rho^{1/3}$ rather than $\epsilon_\infty$ itself. 

The first thing to notice in Fig.~\ref{fig:flow} is that the wind zone, 
characterized by the central white region with $\varv\approx c$ and 
oblate shape, is smaller in run A, because
of the stronger pinching forces around the equator. This difference in the
size of the termination shock in the two cases will have visible consequences 
in the emission maps. The flow structure is similar: in both cases 
substantial velocities are present in the equatorial plane as well as 
along the polar axis (the supersonic jets).
It should be emphasized that the values we find, $\varv\approx 0.5 - 0.8\,c$,
are in agreement with
those observed in the Crab Nebula (Hester et al. \cite{hester02}) and
in Vela (Pavlov et al. \cite{pavlov03}).
The complex flow pattern around the termination shock is also
similar, but hoop stresses, which divert the flow toward the polar axis
through small scale vortices,
occur closer to the termination shock itself in case A, due to a 
much larger region of increased magnetization around the equator. 
This is clearly visible in
the central panels: equipartition is reached or even exceeded
in a large sector around the equator for run A, while this is the case
basically just along the termination shock front for run B, due to 
local compressions of the magnetic field by the flow vortices. 
Notice that the situation described here is very different from
that encountered in standard stationary or self-similar radial models, 
where the nebular plasma reaches equipartition only at large radii. 
This is entirely due to the relaxation of the 1-D approximation, 
and to the complex flow pattern around the termination shock determined 
by the anisotropic wind energy flux.

\begin{figure*}[t]
\centerline{\resizebox{0.8\hsize}{!}{
\includegraphics{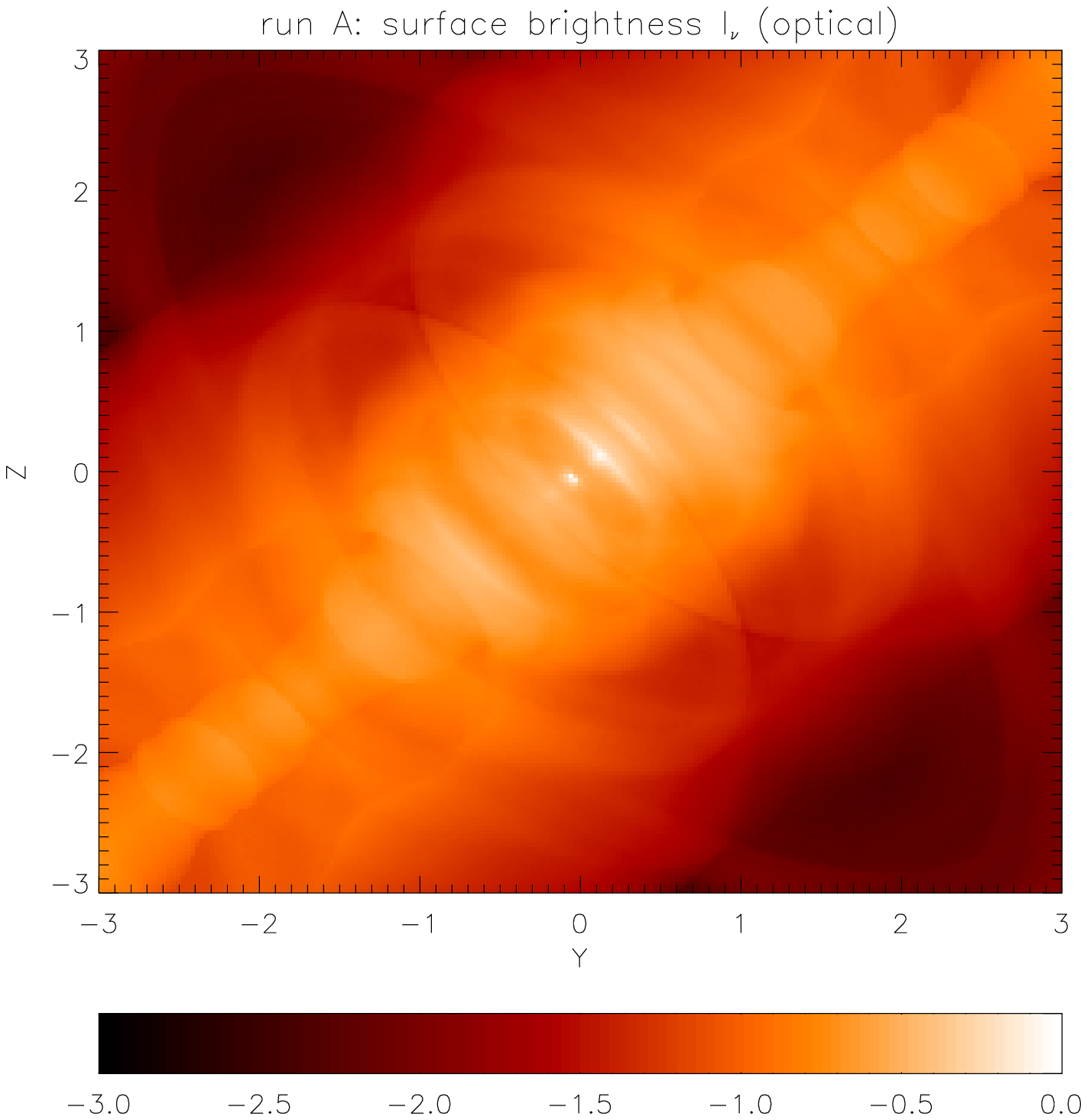}
\includegraphics{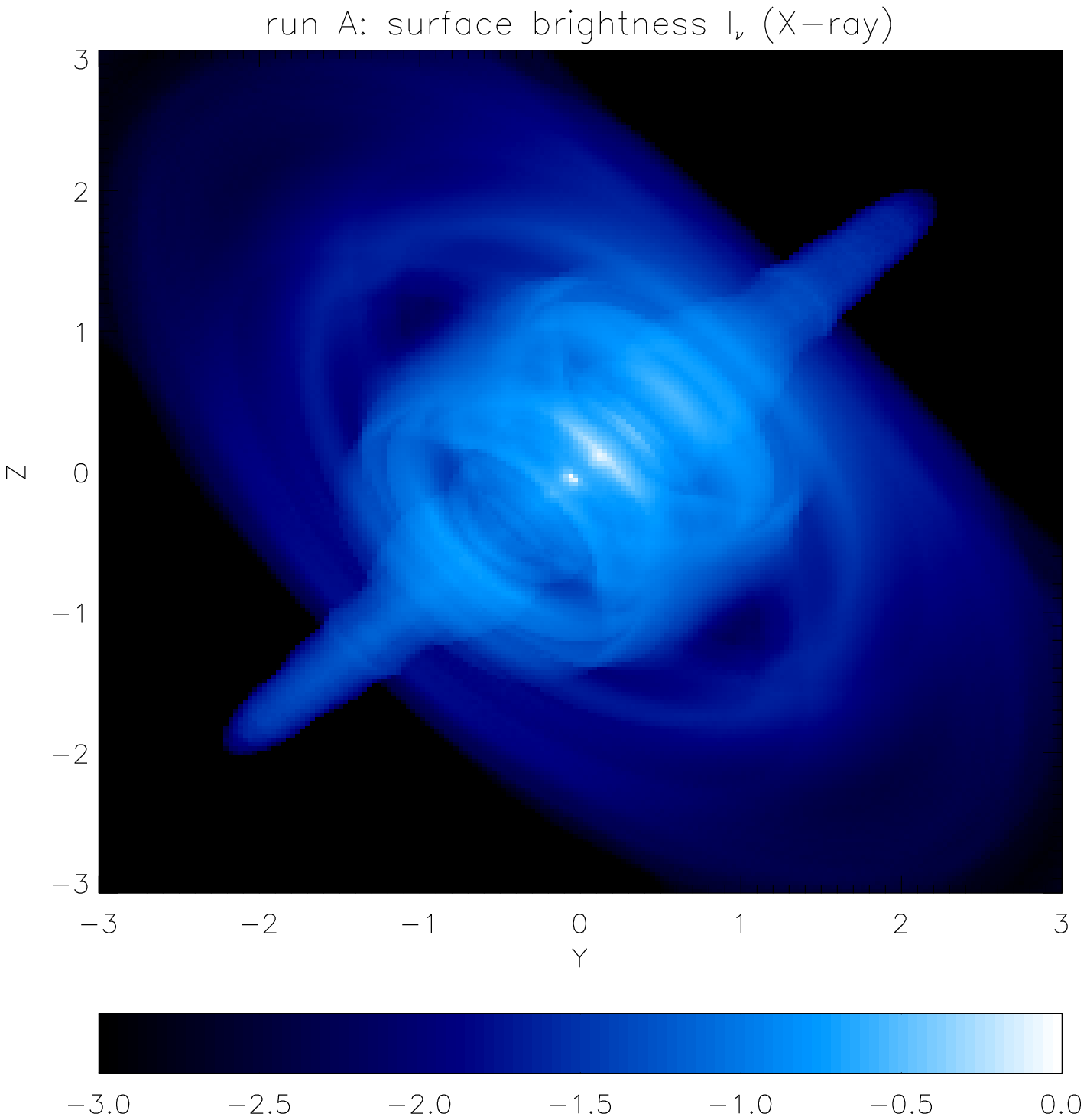}
}}
\centerline{\resizebox{0.8\hsize}{!}{
\includegraphics{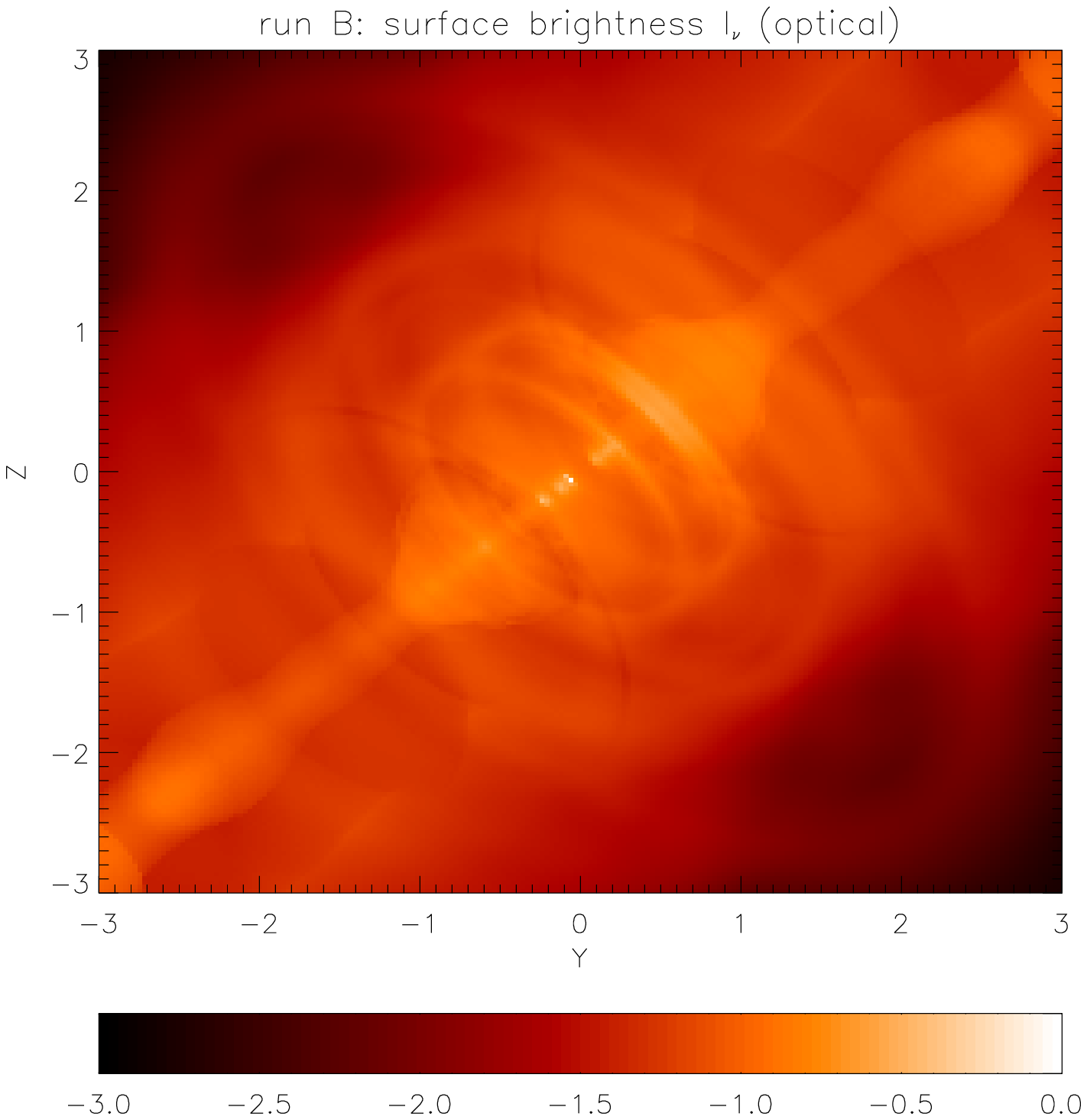}
\includegraphics{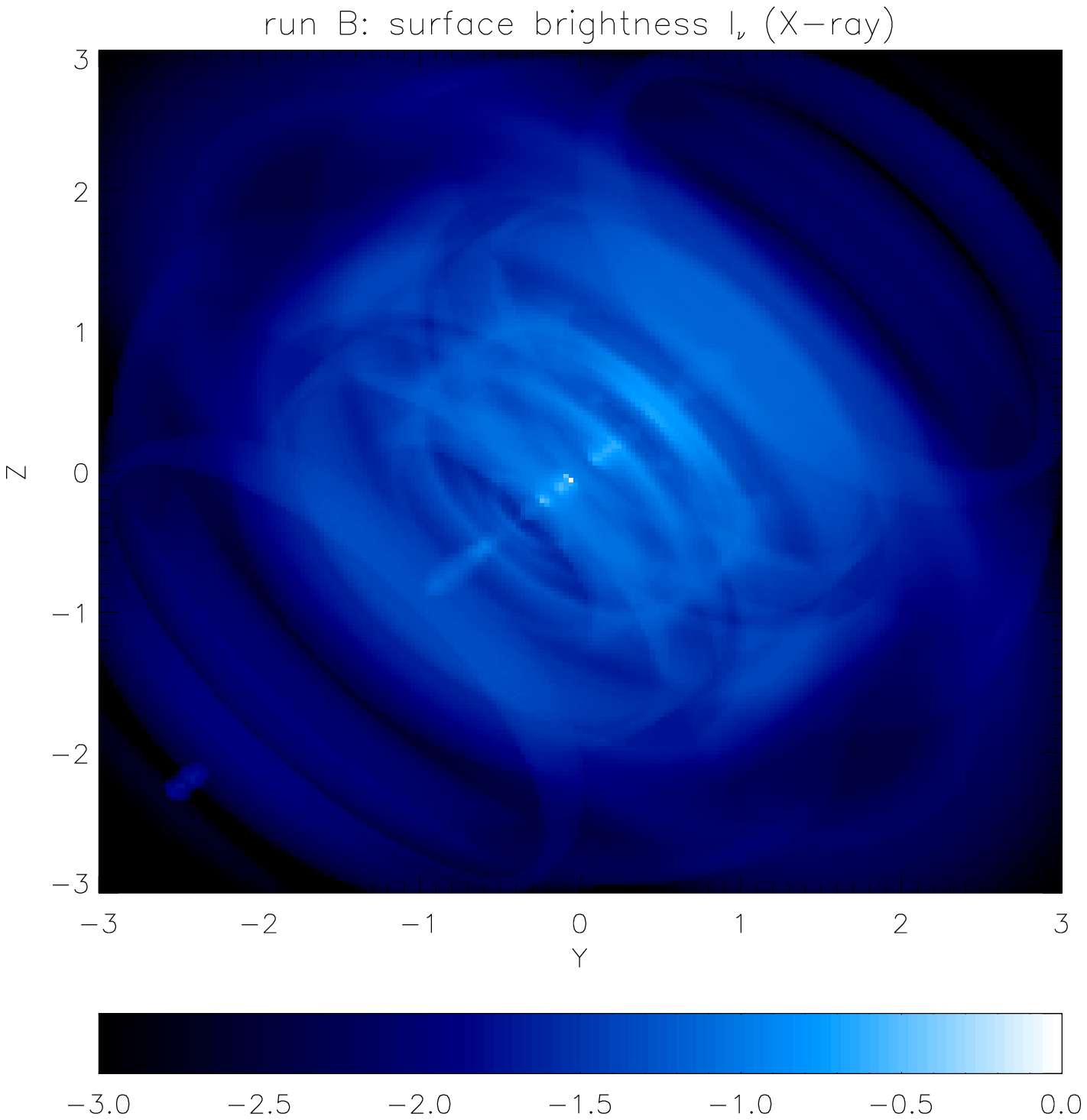}
}}
\caption{
Maps of surface brightness $I_\nu$, in logarithmic scale and normalized 
to the maximum value, for run A (upper row) and run B (lower row). 
Optical images ($5364$~\AA) are displayed in left panels, 
X-ray images ($1$~keV) on the right. The PWN axis of symmetry has
been inclined by $30^\circ$ with respect with the plane of the sky 
and of $48^\circ$ with respect to the North, to compare with images of the
Crab Nebula. For a distance $d\simeq 2$~kpc, we have 1~ly$\simeq 32\arcsec$.
}
\label{fig:brightness}
\end{figure*}

\begin{figure*}[t]
\centerline{\resizebox{0.8\hsize}{!}{
\includegraphics{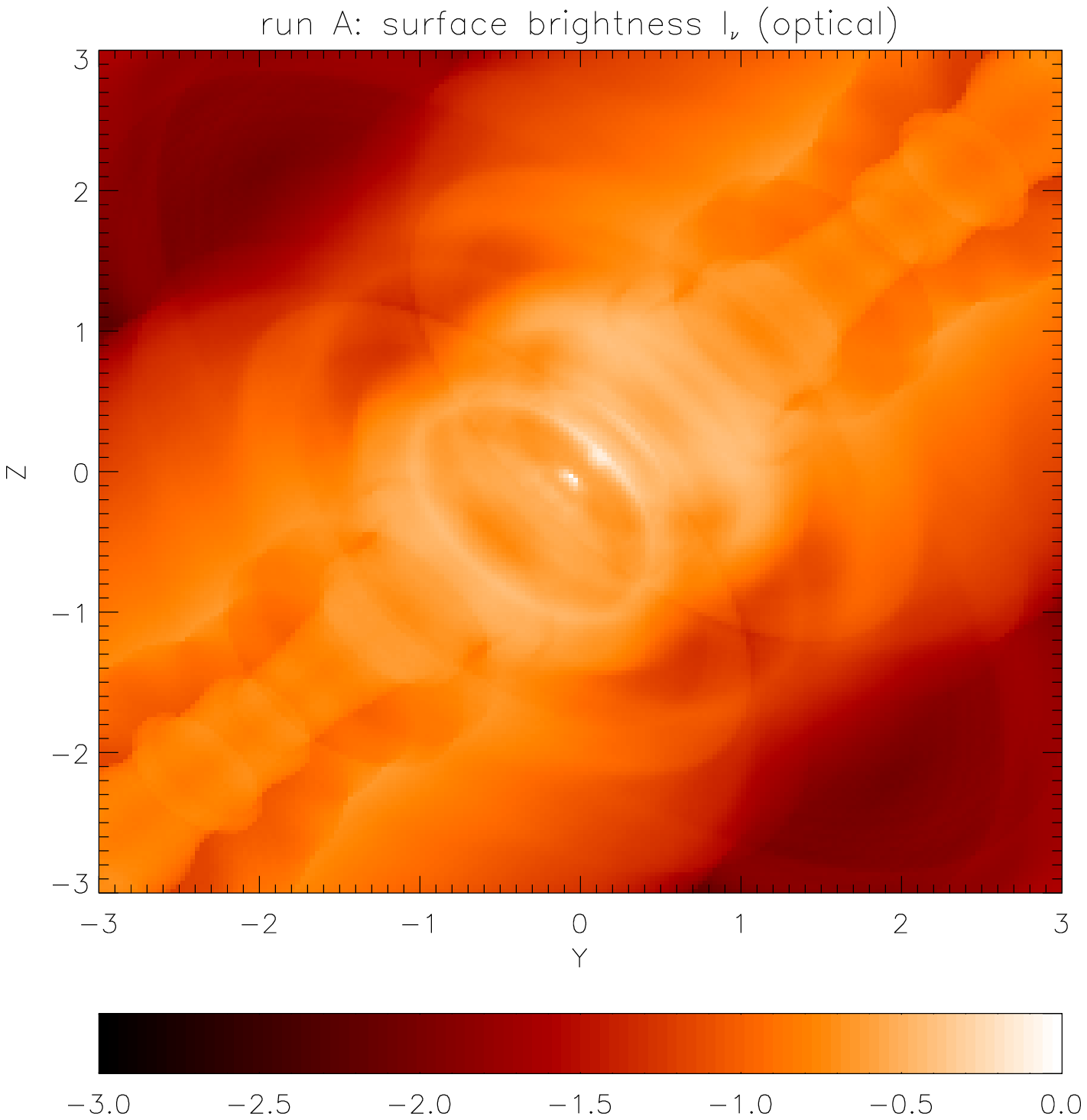}
\includegraphics{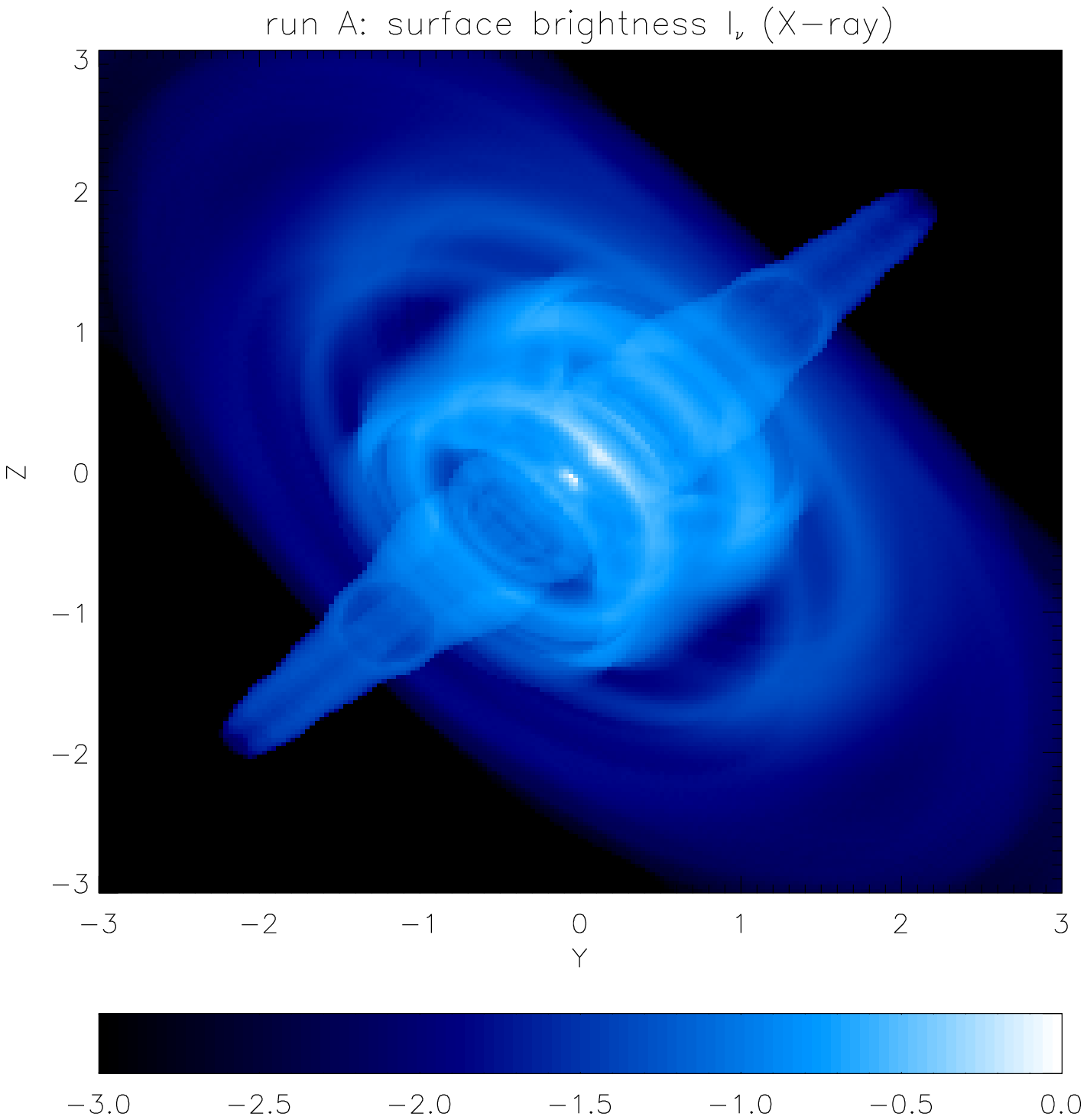}
}}
\caption{Maps of surface brightness for run A corresponding to those
in Fig.~\ref{fig:brightness} obtained with an isotropic magnetic field.
}
\label{fig:brightness2}
\end{figure*}

The two panels on the right
hand side show $\epsilon_\infty$, that is the highest possible 
energy of emitting particles at a given location. As we can see, in both 
cases the pattern resembles that of velocity, as could be expected 
given that the particle energy is advected by the flow. On the other hand, 
these rightmost panels are the ``negative'' of the corresponding central 
ones representing the magnetization structure. This is due to synchrotron 
losses. And again these losses determine the difference between the
maps of $\epsilon_\infty$ for the two runs: while in case A the external 
parts of the map are dark (a lower cut at $\epsilon=10^6$ has been imposed), 
in run B the same regions have a higher maximum energy. This different 
behavior arises from the fact that in run A particles loose most
of their energy in the high magnetization equatorial channel, whereas
in case B, with a lower magnetization there, higher energy particles
survive and are then transported across the nebula by the large scale  
vortices. The structure of the magnetic and velocity fields and the 
resulting distribution of $\epsilon_\infty$ will have important 
consequences, especially for the spectral properties of synchrotron emission, 
as will be widely discussed in Sect.~\ref{sect:res_spectra}.

\begin{figure*}[t]
\centerline{\resizebox{0.8\hsize}{!}{
\includegraphics{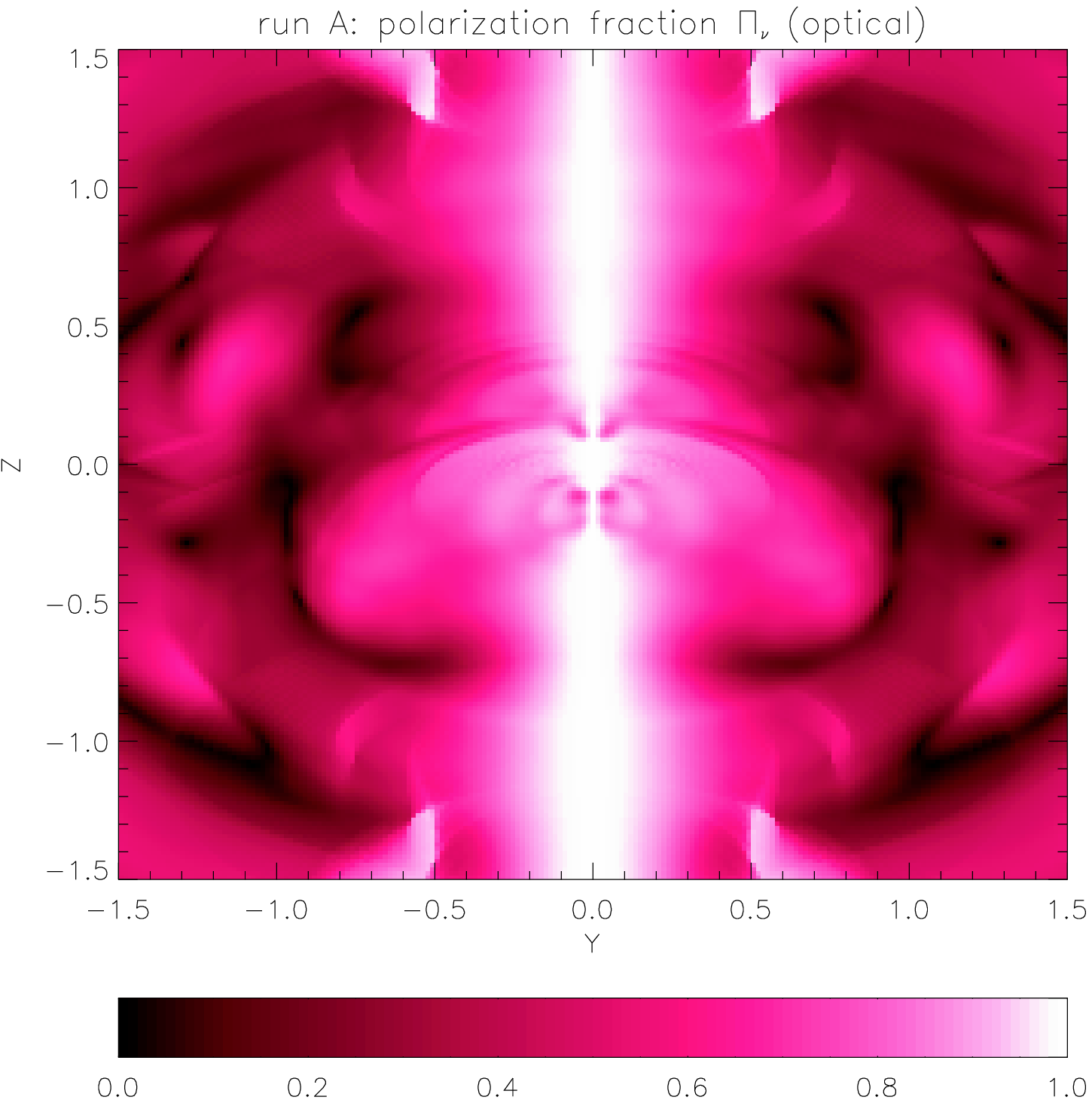}
\includegraphics{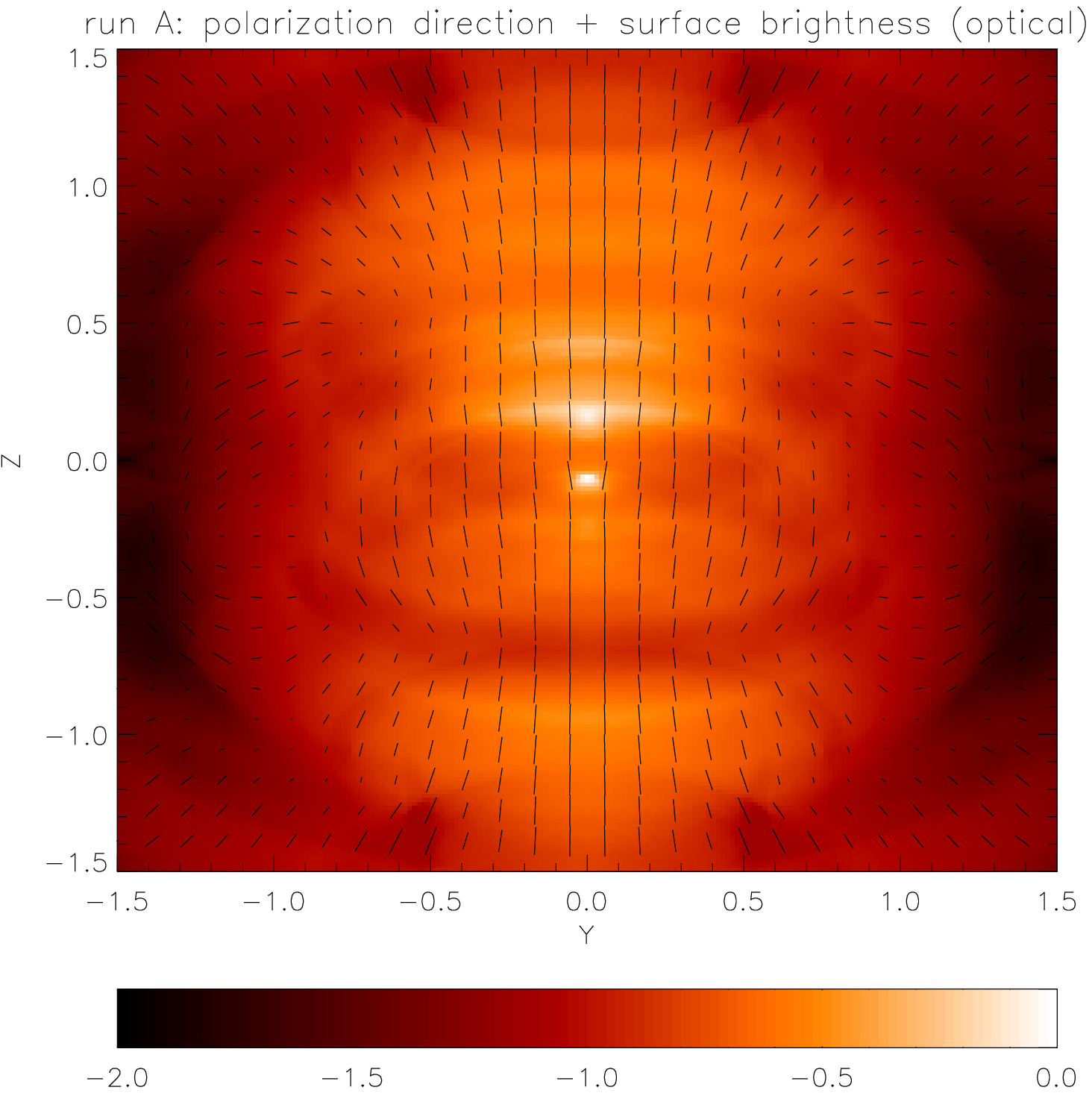}
}}
\centerline{\resizebox{0.8\hsize}{!}{
\includegraphics{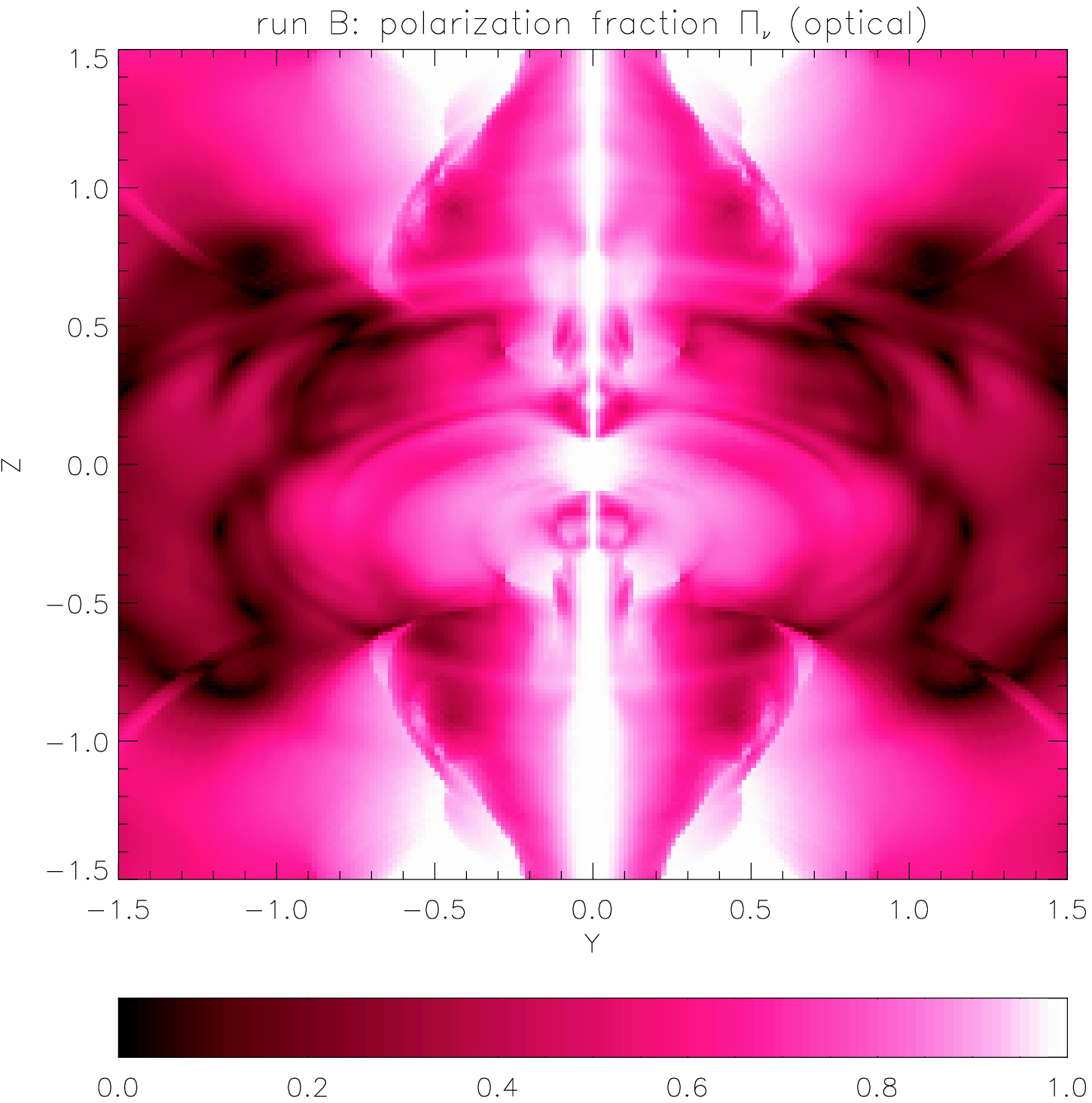}
\includegraphics{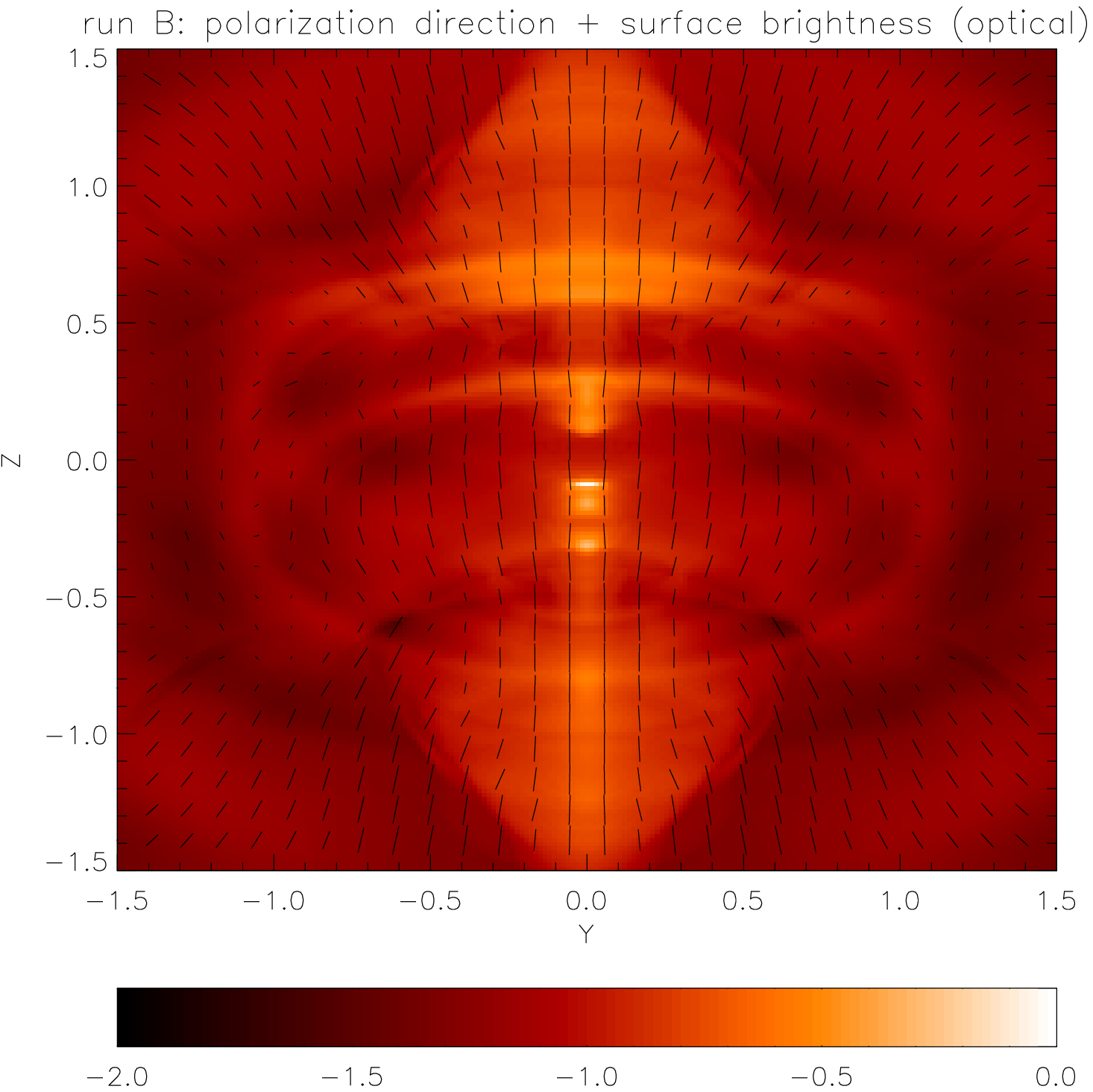}
}}
\caption{
Maps of (optical) polarization fraction $\Pi_\nu$ (left) and direction 
(right, superimposed to the surface brightness map), 
for run A (upper row) and run B (lower row). Polarization fraction is
normalized against $\frac{\alpha+1}{\alpha+5/3}\simeq 70\%$, and the ticks 
length used for polarization direction is proportional to the normalized 
fraction.
}
\label{fig:polarization}
\end{figure*}

\begin{figure*}[t]
\centerline{\resizebox{0.8\hsize}{!}{
\includegraphics{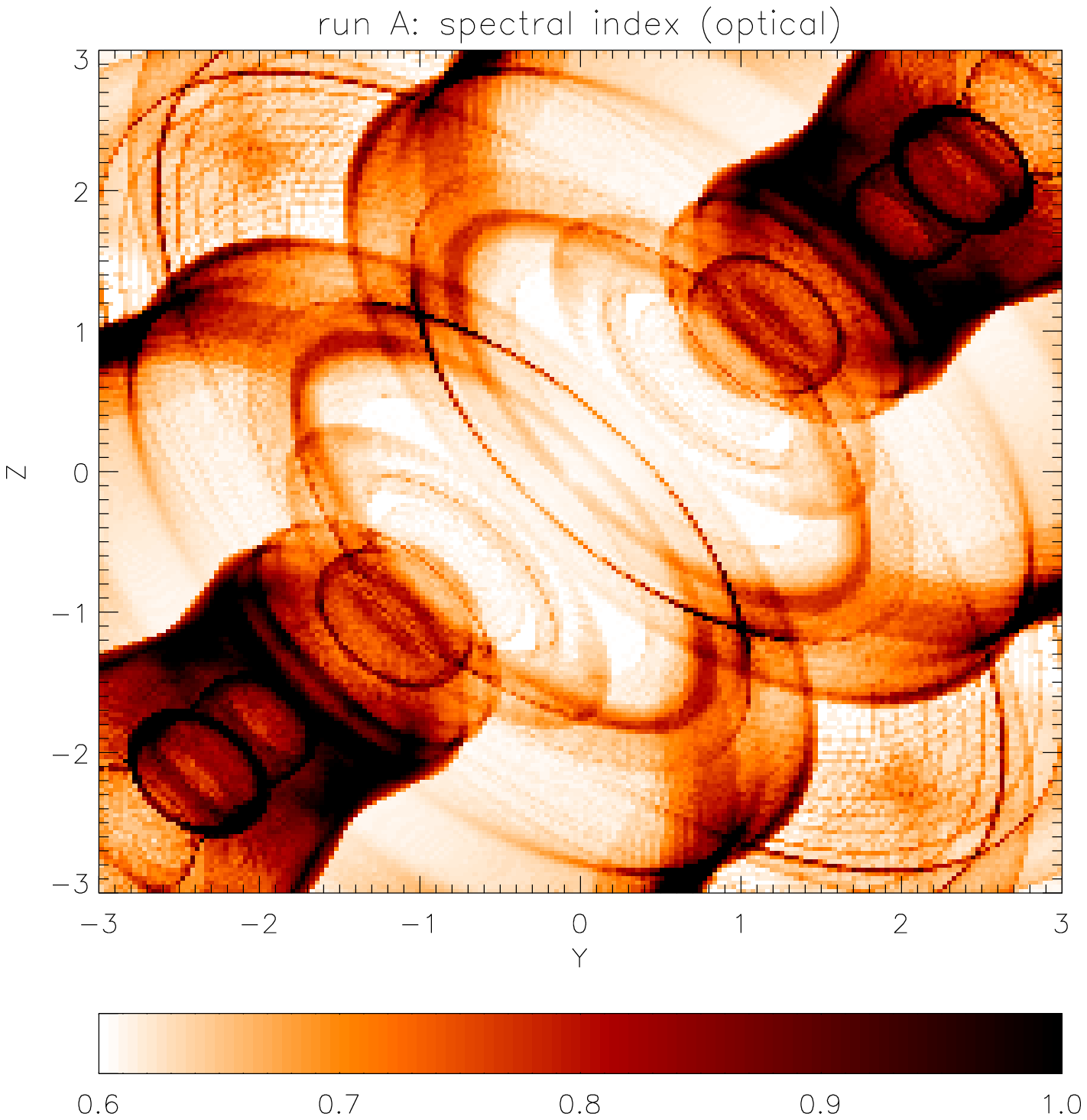}
\includegraphics{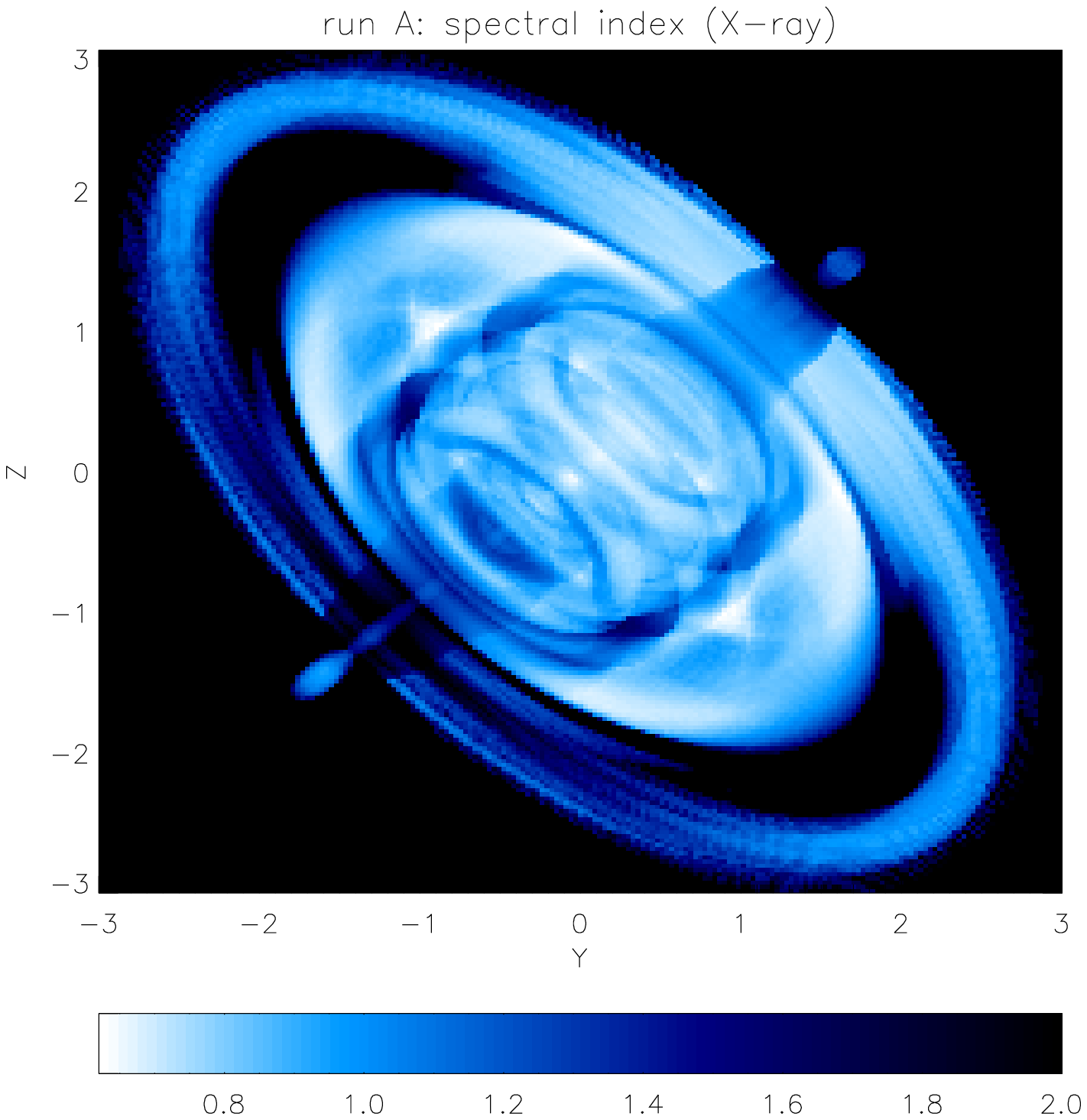}
}}
\centerline{\resizebox{0.8\hsize}{!}{
\includegraphics{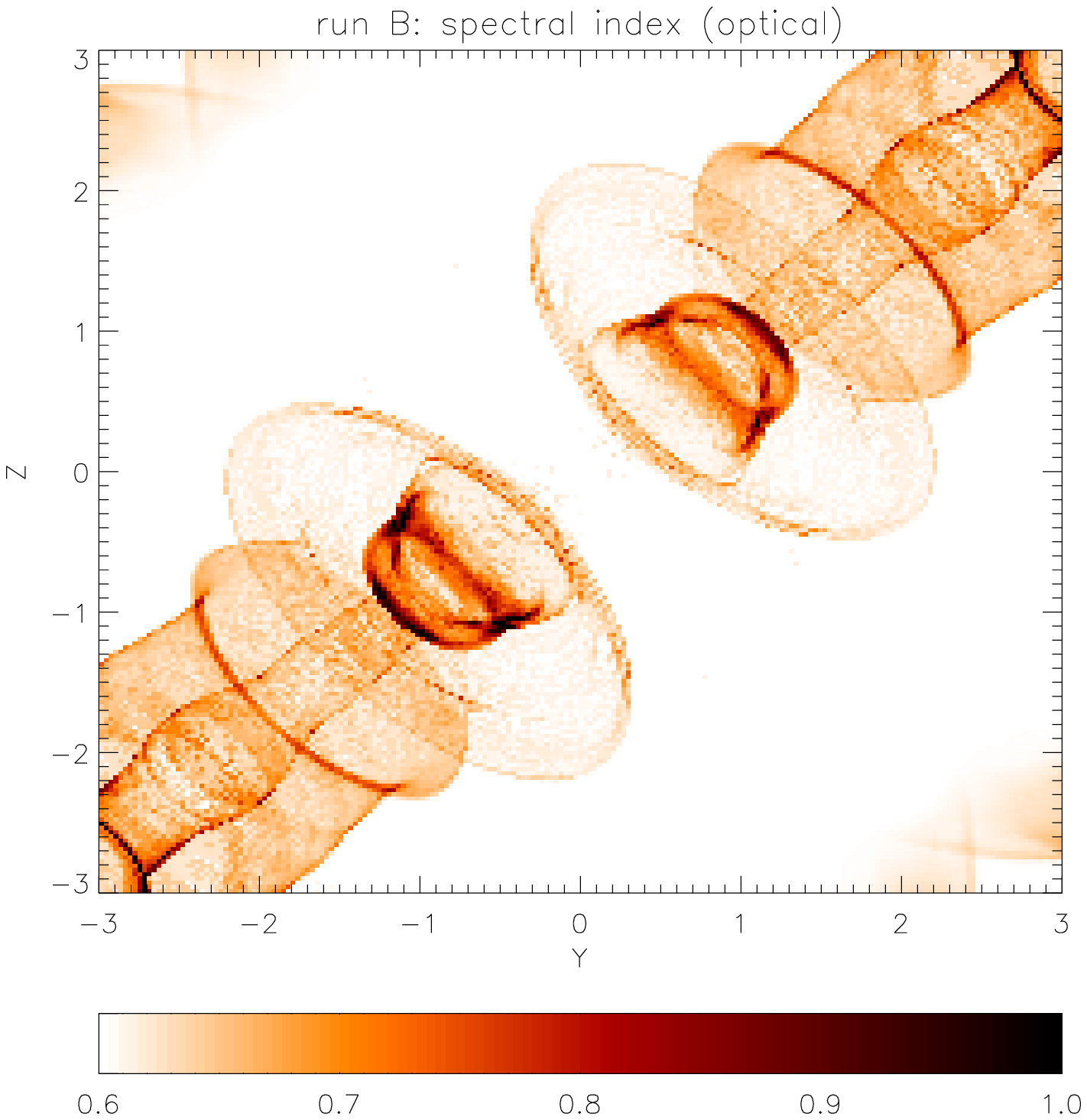}
\includegraphics{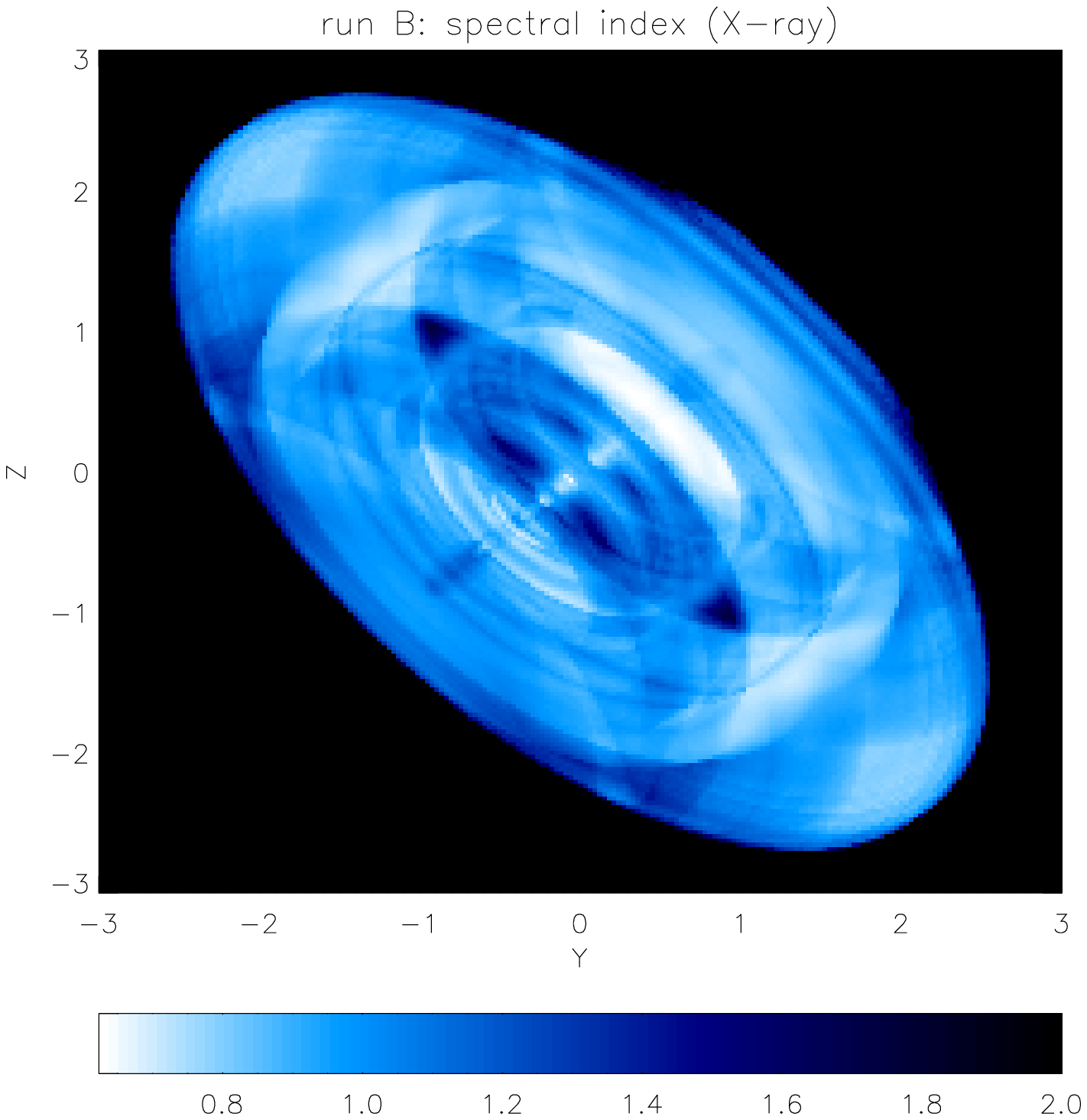}
}}
\caption{
Maps of spectral index $\alpha_\nu$, as defined in Eq.~(\ref{eq:index}), 
for run A (upper row) and run B (lower row). 
Optical images are obtained by comparing specific intensities at
$\lambda=5364$~\AA and $\lambda=9241$~\AA and are displayed in
left panels. X-ray images are obtained by using energies of $0.5$
and $8$~keV, and are displayed on the right.
}
\label{fig:index}
\end{figure*}

\subsection{Surface brightness maps}
\label{sect:res_maps}

In order to obtain synchrotron emission maps, the spectral index 
$\alpha$ of the power law in Eq.~(\ref{eq:f0}) must be assigned. It
is well known that in the case of the Crab Nebula more than a single 
power law at injection is needed in order to explain the observed 
spectrum from the radio to the X-ray band (e.g. Amato et al., \cite{amato00}). 
Here we will assume that particles responsible for optical and X-ray emission
come from the same distribution function, injected at the termination shock 
front, while the radio emitting particles are not considered.
Therefore we take $\alpha=0.6$, in agreement with the values reported
in optical spectral index maps (V\'eron-Cetty \& Woltjer,
\cite{veron-cetty93}) close to the termination shock, 
where synchrotron burn-off has not occurred yet. Note that this
value satisfies the condition $2\alpha-1\ga 0$, needed to approximate
the distribution function in Eq.~(\ref{eq:f_bl}) with the simple power law 
in Eq.~(\ref{eq:f}). We have verified, by measuring the ratios $p/p_0$ and
$p_0/n_0$, that the combined effect of the neglected terms 
leads to changes in the emissivity of $\sim 10\%$, at most.

In Fig.~\ref{fig:brightness} surface brightness maps are shown in optical 
and X-rays for runs A and B (in logarithmic scale and normalized 
to the respective maximum value).
Optical images are calculated for $\lambda=5364$~\AA~ (left panels), one of
the wavelengths selected by V\'eron-Cetty \& Woltjer (\cite{veron-cetty93}),
at which emission from the outer filaments is negligible. For the 
X-rays, instead, we use $h\nu=1$~keV, a value in the range of both 
the satellites \emph{Chandra} and \emph{XMM-Newton}. 
The images are obtained by applying Eq.~(\ref{eq:I}) 
and assuming an inclination of the symmetry axis of $30^\circ$ 
with respect to the plane of the sky and of $48^\circ$ with respect
to the North, values appropriate for the Crab Nebula 
(e.g. Weisskopf \cite{weisskopf00}).
A square box of $6\,\mathrm{ly}\times 6\,\mathrm{ly}$, centered on 
the pulsar position, is used for our synthetic images.

The first thing to notice is that the emission is more distributed
at optical frequencies compared with the X-rays, where the external regions
appear darker. This is the expected effect of synchrotron burn-off,
that causes the sources to appear smaller with increasing observation 
frequency. The central regions are instead rather similar in the two bands,
with evidence for brighter features like rings, arcs and
a central \emph{knot}. The latter, in particular, tends to dominate the 
overall emission (especially in run B) if a linear scale is employed for 
the image.
As it was first shown by Komissarov \& Lyubarsky (\cite{komissarov04}), these
bright features are the result of Doppler boosted emission from fluid
elements moving at relativistic speeds toward the observer. From a 
comparison with Fig.~\ref{fig:flow} it is clear that the regions with high
velocity and magnetization, and thus higher emissivity, are those around
the termination shock: the multiple rings observed are due to the
external vortices produced by the hoop stresses which divert the flow,
the brighter arc is due to flow along the termination shock, while the
knot is due to the fluid escaping from the polar cusp-like region toward
the observer (see the above cited paper for a graphical explanation). 

Morphological differences arise from the comparison of run A and B,
in the X-ray maps especially, where the inner structure is highlighted.
Run A is probably more reminiscent of the Crab Nebula: the external 
torus and the two polar jets are well visible, as well as the knot
(first discovered by Hester et al. \cite{hester95} in optical 
\emph{Hubble Space Telescope} images) and a system of \emph{wisps},
with a brighter arc in what is usually called the inner ring
(Weisskopf et al. \cite{weisskopf00}). Notice that for this feature
we find a radius of 0.18~pc, which is reasonably close to the observed
value of 0.14~pc.
The torus is associated with the higher magnetization equatorial 
region approximately between 1 and 3~ly from the pulsar (see again
Fig.~\ref{fig:flow}), in particular
the emission is higher at $\sim 2$~ly, where the flow converges advecting
emitting particles coming from higher latitudes (see the corresponding
zone with higher values of $\epsilon_\infty$). A feature common to both
structures is the fading of the emission toward the borders, far from 
the polar axis, where $|\vec{B}^\prime\times\vec{n}^\prime|$ becomes 
small in the coefficient of Eq.~(\ref{eq:j}), due to the hypothesis of 
a purely toroidal field (see the discussion below).

The jets are hollow, since the toroidal field vanishes on the axis,
but in run A their emission is significant because emitting
particles are efficiently transported there by the high speed flow.
This last effect is reduced in run B, where the X-ray jets basically
disappear, while the diffuse X-ray emission is enhanced in the nebula. 
The explanation is again found by looking at Fig.~\ref{fig:flow}: 
in run B, the hoop stresses are less efficient, due to the wider low
magnetization region, and particles lose more energy through 
synchrotron radiation before flowing toward the axis.
It is important to notice that, in both cases, jets are visible in the optical
band, and actually, at lower frequencies, the difference between their 
appearance in the two runs is reduced. However they are not so bright now
with respect to their surroundings and the diffuse emission from the outer 
nebula hides them, precisely as it happens in the observations.
In the central region, other than the (overly) bright knot, 
two boosted arcs are visible 
in run B, reminiscent of the situation encountered in the Vela PWN 
(Helfand et al. \cite{helfand01}; Pavlov et al. \cite{pavlov01}). 

Summarizing, it appears that run A, which has a more magnetized wind
near the equator, leads to a brighter X-ray torus and to a single inner 
Doppler boosted arc, whereas run B, where the low magnetization region 
around the equator is more extended, implies a more diffuse nebular 
emission and a more complex system of rings and arcs in the central region.

Before concluding this sub-section, let us comment on the consequences
of the assumption of a purely toroidal magnetic field. Although
evaluation of the dynamical effects of a poloidal field component 
would require a full 3-D treatment, that is beyond our present goal, 
the effects of an isotropic field on the emission properties are
straightforward to check.
If we thus substitute the term $|\vec{B}^\prime\times\vec{n}^\prime|$ 
in the emissivity with its average over the solid angle 
(compare with Eq.~(\ref{eq:b_times_n})):
\be
\label{eq:isotropic}
B^\prime_\perp=|\vec{B}^\prime\times\vec{n}^\prime|=
\sqrt{\frac{2}{3}}\frac{B}{\gamma},
\ee
we obtain the results shown in Fig.~\ref{fig:brightness2}, where
only case A has been considered (similar differences arise for case B).
The main difference, at a given frequency, between the
corresponding images in Fig.~\ref{fig:brightness} and 
\ref{fig:brightness2} is the relative brightness of some of the small
scale features mentioned above. First of all the knot appears to be
less dominant than before, as well as the main arc in the inner ring.
Moreover, as one would expect, the outer parts of all ring-like structures
are enhanced. In particular, looking at the X-ray map, we notice that
what we here have referred to as the inner ring is now more uniform, 
as in the observations (Weisskopf et al. \cite{weisskopf00}).

\subsection{Polarization maps}
\label{sect:res_pol}

Let us now concentrate on the (linear) polarization properties of the
nebular emission. A preliminary study was reported in an earlier paper
(Bucciantini et al. \cite{bucciantini05b}), where maps of the polarization
fraction ($\Pi_\nu$) and direction based on a similar simulation were 
presented.
In Fig.~\ref{fig:polarization} we show the same maps for runs A and B,
where the central region is zoomed and not rotated in the plane of the
sky for ease of interpretation (the tilt is retained).
The two cases are similar: the polarized fraction is high along the polar
axis, where the projected magnetic field is always orthogonal to the line
of sight,
while depolarization occurs in the outer regions, where contributions
from projected fields with opposite signs sum up along the line of sight.
On the right hand side 
we show the polarization direction, with ticks proportional
to the fraction displayed on the left, together with the corresponding map
of surface brightness, with the aim of making clear the association
between the polarization behaviour and the emission
main features. As we can see, polarization ticks are basically always
orthogonal to the toroidal field, displaying the behaviour expected
given the inclination of the symmetry axis with respect to the plane of the
sky. However, in the rings where the velocity is relativistic, deviations
of the vector direction due to polarization angle swing are also
visible (e.g. in the arc of the torus, that is the second brightest arc 
above the central knot).

Concerning differences between the two runs, we notice that the structure
of polarization fraction and direction becomes more complex for run B, 
for which also the magnetization map in Fig.~\ref{fig:flow} was more 
complicated, leading to the presence of multiple rings. 
Notice also that in these zoomed maps it is possible
to see a small displacement (to the South) of the knot with respect to 
the central position, as in the figures by Hester et al. (\cite{hester95}).
Unfortunately, high resolution optical polarization maps of the inner
region of the Crab Nebula are not available yet: all papers mentioned
in the Introduction refer to the whole nebula, where contributions
from the confining medium may be significant. Finally, for X-ray
polarimetry, which could really provide crucial clues to the magnetic
structure in the inner region, the necessary technology is still to come.

\subsection{Spectral index maps and integrated spectra}
\label{sect:res_spectra}

\begin{figure}[t]
\vspace{4mm}
\resizebox{\hsize}{!}{\includegraphics{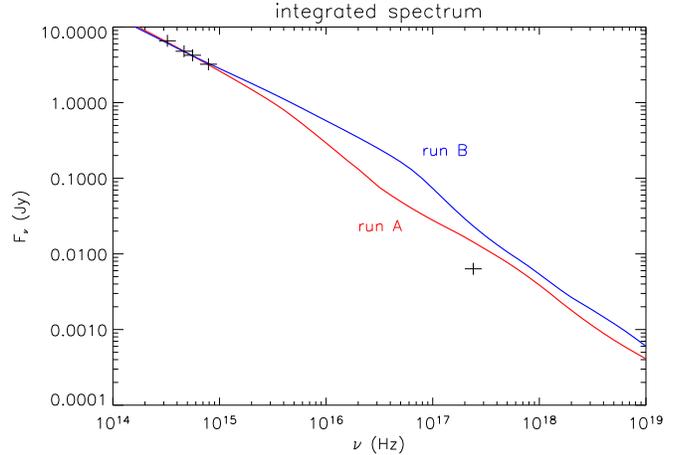}}
\caption{The integrated spectrum as a function of frequency for the
two runs. Spectra are normalized to $\lambda=5364$~\AA.}
\label{fig:spectrum}
\end{figure}

Consider now the spectral properties, starting with the maps of
spectral index $\alpha_\nu$ in Fig.~\ref{fig:index}, both in the optical 
and X-ray bands.
Optical images are obtained by comparing specific intensities at
$\lambda=5364$~\AA~ and $\lambda=9241$~\AA, wavelengths that have
been chosen so as to allow direct comparison with the observational data
presented in the paper by V\'eron-Cetty \& Woltjer (\cite{veron-cetty93}). 
The X-ray spectral index maps, instead, are obtained by using energies 
of $0.5$ and $8$~keV, as in Mori et al. (\cite{mori04}).

The optical images do not show significant variations from the
injection value of $\alpha=0.6$ in the inner region. Once again run A
seems to be more closely representative of the situation in the Crab Nebula,
with spectral softening occurring, starting from a distance of about 0.5 
pc from the central pulsar position, in the vicinity of the polar axis, 
and more gradually at larger cylindrical radii. This is consistent with 
observations, at least qualitatively, since softening, in the optical band,
is present only beyond the torus. A general consideration is that in our runs, 
and especially in run B, the nebular magnetic field in the 
central region seems to be weaker than the value usually 
assumed for the average field in the Crab Nebula $\sim 3\times 10^{-4}$~G.
In the torus, our simulations give a field strength lower by a factor
of 2, for run A, and 3, for run B. In addition to that,
when comparing to real images, one should also consider that dimensions 
must be rescaled somehow. While in run A the positions of the main features 
are roughly in agreement with the observations (see the discussion in
Sect.~\ref{sect:res_maps}), in run B the radii of the inner ring and the
torus are approximately two times larger than observed. 
This naturally concurs to the spectral flatness of the corresponding map. 

In the X-rays synchrotron burn-off occurs on much shorter distances
and the maps are much more shaped: spectral softening is much stronger and, 
as a consequence, basically only the inner structure
is visible in the adopted range. Notice how the values are similar to those 
measured by Mori et al. (\cite{mori04}) (where photon index is used for 
the scale, that is $\alpha_\nu +1$), and in run A we even observe the 
brighter (harder) features in the polar jets, especially the southern one.

Finally, let us calculate the integrated spectra (net fluxes) as in 
Eq.~(\ref{eq:netflux}), for the two runs A and B, and compare them with 
spectro-photometric data of the Crab Nebula, at optical and X-ray frequencies. 
In Fig.~\ref{fig:spectrum} the net flux $F_\nu$ (in Jansky) is plotted 
against frequency $\nu$ (in Hertz). Solid lines are the synthetic integrated 
spectra obtained from simulations. The four crosses in the optical range 
correspond to data at $\lambda=9241$, $6450$, $5364$, and $3808$~\AA, 
corrected for interstellar absorption and thermal contribution from the 
filaments (V\'eron-Cetty \& Woltjer \cite{veron-cetty93}), whereas the X-ray
data point at 1~keV is taken from Willingale et al. (\cite{willingale01})
and refers to \emph{XMM-Newton} observations.
We can notice the different behaviour of the two runs. For run A
the spectrum steepens because of synchrotron burn-off already in 
the optical band, where the slope is 0.75 rather than the injection value 
of 0.6. A spectral break is seen at $\nu_{b1}\approx 2-3\times 10^{15}$~Hz, 
as expected (e.g. V\'eron-Cetty \& Woltjer \cite{veron-cetty93}), with
the slope changing to 1.15, yielding an increase of 0.55, comparable 
to the value 0.5 usually adopted. On the other hand, in run B, the lower
magnetic field causes losses to be negligible in the optical and the
break to move toward X-ray photon energies. 

The most puzzling feature is the presence in both cases of further,
unexpected, changes of slope, with the appearance of what we term as 
\emph{inverse breaks}, i.e. flattenings of the spectrum.
Focusing on the results relative to run A, we see that the spectral 
index has a second change from 1.15 to 0.8 at around 
$\nu_{b2}\sim 3 \times 10^{16}$~Hz; then it increases again to 1.1 at 
$\nu_{b3}\sim 10^{18}$~Hz; and finally decreases to 0.85 at 
$\nu_{b4}\sim 3 \times 10^{18}$~Hz. The behavior of run B is qualitatively 
analogous, although there are differences in the position and entity of the
changes of slope.

\begin{figure}[t]
\resizebox{\hsize}{!}{\includegraphics{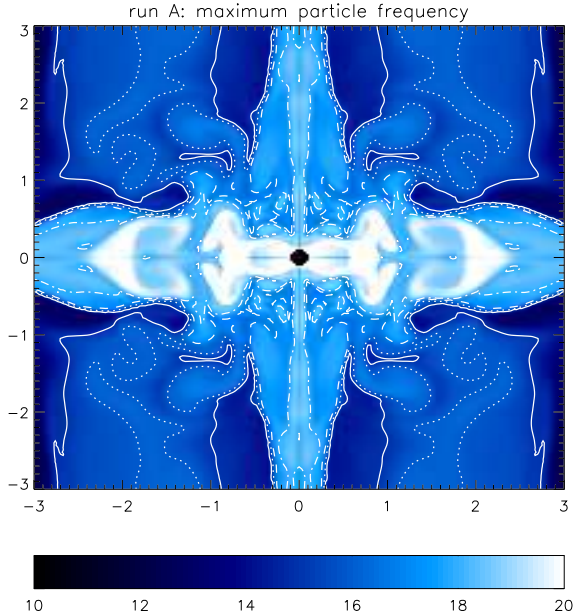}}
\caption{The distribution of the maximum particle frequency (in Hz units), 
that is the critical frequency of Eq.~(\ref{eq:nu_infty}) calculated ignoring
line of sight effects. Contours refer to the values of $10^{15}$~Hz
(solid line), $10^{16}$~Hz (dotted line), $10^{17}$~Hz (dashed line), 
$10^{18}$~Hz (dot-dashed line).}
\label{fig:nu_infty}
\end{figure}

The reason for this behavior is better understood by looking at 
Fig.~\ref{fig:nu_infty} where we show the cut-off 
frequency for run A as a function of space. In particular, if one 
looks at the contours, it is clear that while the initial break 
at $\nu_{b1}$ is due to loss of emission from high latitudes, that
particles with energies corresponding to emission at $\nu>\nu_{b1}$
are not able to reach (consider the contours referring to 
$10^{15}$, solid line, and $10^{16}$~Hz, dotted line, in the figure), 
the \emph{inverse break} at $\nu_{b2}$
appears because there is a range of frequencies for which the emitting 
region keeps constant in size. This is strictly related to the particular 
configuration of the magnetic field (see the magnetization map in 
Fig.~\ref{fig:flow}) that is mainly confined to the equatorial region
due to the presence of large scale vortices.
This is confirmed by the spatial ditribution of
the maximum particle energy $\epsilon_\infty$, which for high frequencies
(thus in X-rays) appears to be concentrated just around the termination
shock, so that a further increase in frequency does not lead to any
further reduction of the emitting region.

The result of the deviations from the expected monotonic spectral 
steepening described above is that X-ray emission is too high, by a factor 
of 2 in run A and 3 in run B.
This difference of values between the two runs mostly depends on 
the different strength of the magnetic field which causes stronger 
synchrotron losses for run A than for run B, although still not enough to 
explain the observations quantitatively. 

\section{Conclusions}
\label{sect:conclusions}

In this paper a complete suite of diagnostic tools for deriving
synchrotron emission properties from numerical relativistic MHD
simulations has been presented. The method is quite simple and
general, so that it can be applied to any numerical scheme and
to any astrophysical source of synchrotron radiation.
The application presented here concerns the emission from PWNe,
based on the axisymmetric model by Del Zanna et al. (\cite{delzanna04}).
In this case, synchrotron radiation is the primary diagnostic for the
physics in the source and may give important clues about the structure 
of the pulsar wind itself, since its properties are found to be 
reflected in the nebular emission. In particular we have studied the
differences arising from two different magnetic structures 
of the wind: 
one in which the striped (low magnetization) wind region is very narrow, 
and the toroidal field has a maximum
near the equator (run A), and another where the maximum occurs at a
much higher latitude above the equator (run B). In both runs the average
magnetization was the same, $\sigma_\mathrm{eff}\approx 0.02$, and high
enough to induce the presence of supersonic polar jets.

The width of the striped wind region of low
magnetization is usually considered as directly linked to the inclination
angle between the pulsar's rotation and magnetospheric axes, thus such
a study could yield crucial information otherwise difficult to obtain.
By looking at surface brightness maps, especially in X-ray, it
appears that run A produces images which are remarkably similar
to that of the Crab Nebula observed by the \emph{ROSAT}
(Hester et al. \cite{hester95}), \emph{Chandra}
(Weisskopf et al. \cite{weisskopf00}), and \emph{XMM-Newton}
(Willingale et al. \cite{willingale01}) satellites. An equatorial
torus and two polar jets are clearly visible, together with an inner
ring, with Doppler boosted emission concentrated in a main arc, and a 
bright knot just SE from the pulsar position. Particularly noticeable
is the presence of the jets, necessarily hollow because of our 
assumption of axisymmetry and of a purely toroidal field (though
the case of an isotropic field has been treated here too):
these did not appear as emitting features in previous simulations. 
Run B produces instead maps which resemble vaguely the Vela PWN 
(Helfand et al. \cite{helfand01}; Pavlov et al. \cite{pavlov01}, 
\cite{pavlov03}),
with a couple of bright arcs NW from the central knot in the inner
region. This is entirely due to the different flow structure determined
by the lower magnetization around the equator (hoop stresses and
small scale vortices develop at higher latitudes).

In the present paper we have used observations mainly of the Crab 
Nebula as a benchmark for our diagnostic techniques, for the simple reason
that the Crab Nebula is by far the best studied PWN. We have briefly mentioned
the Vela case too, but one may wonder whether by tuning our model parameters 
the resulting surface brighness maps could reproduce also different features 
encountered there or in some other PWN. One of the most puzzling problems
has always been the jet in the Vela PWN, which is much brigher 
on the \emph{wrong} side of the pulsar position (that for which the jet 
is receding from us). The best explanation is probably to invoke kink-type 
instabilities that manage to bend sections of the counter-jet toward us,
thus enhancing their emission, but in order to test this idea full 3-D 
simulations would be required.
Regarding other objects, we may consider the case of
the PWN around pulsar B1509-58. In the central region multiple knots are 
observed (Gaensler et al. \cite{gaensler02}) and the brightest ones are 
located between the pulsar position and the boosted arc on the torus, 
opposite to the direction of the (main) jet. 
Again, these features cannot be explained within
our 2-D model, possibly they could arise due to time dependent small-scale
Kelvin-Helmholtz instabilities, as it has been suggested for the
variable optical \emph{wisps} and \emph{sprites} in the Crab Nebula
(Begelman \cite{begelman99}; Bucciantini \& Del Zanna \cite{bucciantini06}),
but in any case the approximation of axisymmetry must be relaxed.
In the same source, the ring-like region named RCW 89 could be produced by the
counter-jet (very faint in the \emph{Chandra} images) hitting the contact
discontinuity between the PWN and the expanding ejecta. This scenario is
well within our assumptions and its accurate modeling is left as future work.

We have also computed detailed polarization maps. This is
expected to be a very powerful diagnostic technique for the 
magnetic structure of the inner regions of PWNe, since synthetic 
surface brightness maps are likely to be too largely dominated by 
Doppler boosted features, in the vicinities of the central source, 
to be sensitive to small changes in the magnetic field strength 
and direction 
(see Bucciantini et al. \cite{bucciantini05b} for a wider discussion).
The results show a high degree of linear polarization 
close to the axis of symmetry (not far from the theoretical maximum of 
$70\%$, given the spectral index $\alpha=0.6$), and a strong depolarization
in the outer equatorial regions. Relativistic effects related to
Doppler boosting (polarization angle swing) are also visible in
the brighter arcs. High resolution optical, not to mention X-rays,
polarization observations, needed for a comparison of the numerical 
results with the real data, aimed at probing the inner PWN magnetic 
structure, are lacking, at present, but are highly desirable given 
their diagnostic potential. 

Synthetic spectral index maps have been computed here for the first
time on the basis of MHD numerical simulations. At a qualitative
level, the results of our run A are in good agreement with both optical
(V\'eron-Cetty \& Woltjer \cite{veron-cetty93}) and X-ray (Mori et al. 
\cite{mori04}) data: spectral softening is weak in the central
region and stronger at the borders of the torus, especially in the 
X-rays, where even the harder features inside the jets
highlighted in {\em Chandra} data are reproduced.
However, the nebular field appears to be too low to explain the
observations quantitatively, giving too weak a softening, especially
for run B. This is further confirmed when looking at the integrated spectra, 
which show that the spectral break is too weak and occurs at a too high 
frequency. The result is that, if we normalize the spectra to a flux in the 
optical, X-ray data are over-estimated by a factor 2 for run A and 3 for run B.
We have attributed this behavior to the distribution of the nebular
magnetic field and, consequently, to that of the maximum energy of
emitting particles which serves as a local synchrotron cut-off.
Such distributions basically show that the magnetic field is too 
strongly affected by the compression due to large scale 
flow vortices and piles up around the termination shock.
We have also verified that this situation invariably occurs for 
different values of the simulation parameters. 

In spite of these successes in PWN modeling, in order to proceed further 
3-D simulations are certainly required. The strongest limitation
appears to be the assumption of a purely toroidal magnetic field,
both for the overall dynamics (e.g. allowing for kink-type instabilities 
along the axis and possibly reconnection at the equator) and for the
emission properties, where a certain amount of disordered magnetic
field on small scales could help to reproduce the observations, as we
have shown in maps for the optical and X-ray bands in the case of an
isotropic field (see also the discussion in Shibata et al. (\cite{shibata03}). 
Another issue not properly addressed here is time variability:
though the overall evolution of the PWN / SNR system is nearly
self-similar in the free expansion phase, the small scale vortices 
around the termination shock are variable on a time scale of a few years
(see also Bogovalov et al. \cite{bogovalov05}).
We leave the discussion of these two important issues to future papers. 

\begin{acknowledgements}
The authors thank R. Bandiera, F. Pacini, J. Arons, S. Shibata, and
S. Komissarov for valuable discussions, and the anonymous referee 
for helping us to improve the paper.
This work was partly supported by MIUR under grant Cofin 2004 (Pacini).
\end{acknowledgements}

\end{document}